\documentclass[traditabstract]{aa} % for the abstract without structuration 
\usepackage{graphicx}
\usepackage{color}
\usepackage{txfonts}
\usepackage{natbib}
\bibpunct{(}{)}{;}{a}{}{,}

\newcommand{\ltsima}{$\; \buildrel < \over \sim \;$}
\newcommand{\simlt}{\lower.5ex\hbox{\ltsima}}
\newcommand{\gtsima}{$\; \buildrel > \over \sim \;$}
\newcommand{\simgt}{\lower.5ex\hbox{\gtsima}}

\newcommand{\be}{\begin {equation}}
\newcommand{\ee}{\end {equation}}

\begin{document}
\title{Hydrodynamic capabilities of an SPH code incorporating an
artificial conductivity term with a gravity-based signal velocity}

   \subtitle{}

   \author{R. Valdarnini
          \inst{1}
          }

   \institute{SISSA/ISAS, via Bonomea 265, I-34136 Trieste, Italy\\
              \email{valda@sissa.it}
             }

   \date{Received ; accepted }

% \abstract{}{}{}{}{} 
% 5 {} token are mandatory
 
  \abstract{
{This paper investigates the hydrodynamic performances of an SPH code 
incorporating an artificial heat conductivity term in which the adopted signal
velocity is applicable when gravity is present.
To this end, we analyze results from simulations produced using a suite of
standard hydrodynamical test problems.}
{In accordance with previous findings it is shown that the performances of 
SPH to describe the development of Kelvin-Helmholtz instabilities depend
strongly on the consistency of the initial condition set-up and  
on the leading error in the momentum equation due to incomplete kernel sampling.
On the contrary, the presence of artificial  conductivity does not
significantly affect the results.} 

{An error and stability analysis shows that 
the quartic $B-$spline kernel ($M_5$) possesses very good stability properties 
and we propose its use with a large neighbor number, between $\sim50$ (2D) to 
$\sim 100$ (3D), to improve convergence in simulation results without being 
affected by the so-called clumping instability.}
{Moreover, the results of the Sod shock tube test demonstrate that in order 
to obtain simulation profiles in accord with the
analytic solution, for simulations employing kernels with a  non-zero 
first derivative at the origin it is necessary the use of a much larger 
number of neighbors than  in the case of the $M_5$ runs.}

{SPH simulations of the blob test show that in order to achieve blob disruption
it is necessary the presence of an  artificial conductivity  term.
However, it is found that in the regime of strong supersonic flows 
an appropriate limiting condition, which depends on the Prandtl number, 
must be imposed on the artificial conductivity  SPH coefficients in
order to avoid an unphysical amount of heat diffusion.}

{Results from hydrodynamic  simulations that include self-gravity show profiles 
of hydrodynamic  variables that are in much better agreement with those 
produced using mesh-based codes. In particular, the final levels of core 
entropies in cosmological simulations of galaxy clusters are consistent 
with those  found  using AMR codes. This demonstrate that the proposed 
 diffusion scheme is capable of mimicking the  process of 
entropy mixing which is produced during structure formation because of
diffusion due to turbulence.}

{Finally, results of the Rayleigh-Taylor instability  test demonstrate that 
in the regime of very subsonic flows the code has still several difficulties
 in the treatment of hydrodynamic  instabilities. These problems being  
intrinsically due to the way in which in standard SPH gradients are calculated 
and  not to the implementation of the artificial conductivity  term.
To overcome these difficulties  several numerical schemes have been 
proposed which, if coupled with the SPH implementation presented in this 
paper, could solve the issues  which recently have been addressed in 
investigating SPH performances to model subsonic turbulence.
}
}

%   \keywords{Hydrodynamics: SPH: artificial conductivity -- Methods: numerical

%          }


\titlerunning{ Hydrodynamic  of an SPH code incorporating an
artificial conductivity term }

   \maketitle
%
%________________________________________________________________

\section{INTRODUCTION}
\label{intro.sec}
Smoothed particle hydrodynamics (SPH) is a Lagrangian, mesh-free, 
particle method
which is used to model fluid hydrodynamics in numerical simulations. The technique
was originally developed in an astrophysical context \citep{lu77,gm77}, but
since then it has been widely applied in many other areas \citep{mo05} of 
computational fluid dynamics. 

The method has several properties which make its use particularly 
advantageous in astrophysical problems \citep{hk89,ro09,sp10}. Because of its 
Lagrangian nature the development of large matter concentrations
  in collapse problems is followed naturally, moreover the method is free of 
advection errors and is Galilean invariant,
Finally, the method naturally incorporates self-gravity and possess very good 
conservation properties, its fluid equations being derived from variational 
principles. 

The other computational fluid dynamical method commonly employed
in numerical astrophysics is the standard Eulerian grid based approach in which 
the fluid is evolved on a discretized mesh \citep{st92,ry93,no99b,fr00,te02}. 
The spatial resolution of an Eulerian scheme based
on a fixed Cartesian grid is often insufficient however to adequately resolve the
large dynamic range frequently encountered in many astrophysical problems, such as 
galaxy formation. This has motivated the development of adaptative mesh refinement 
(AMR) methods, in which the spatial resolution of the grid is locally refined 
according to some selection criterion \citep{be89,kr97,no05}.
Additionally, the order of the numerical scheme has been improved by adopting 
the  parabolic piecewise method (PPM) of \cite{co84}, such as in the 
AMR Eulerian codes ENZO \citep{no99b} and FLASH \citep{fr00}.

Application of these different types of hydrodynamical codes to the same 
test problem with
identical initial conditions should in principle lead to similar results.
However, there has been  growing evidence over recent years that in a variety
of hydrodynamical test cases there are significant differences between the 
results produced the two types of methods 
\citep{os05,ag07,wa08,ta08,mi09,read10,vrd10,ju10}.

For instance, \cite{ag07} showed that in the standard SPH formulations the growth 
of  Kelvin-Helmholtz (KH) instabilities in shear flows is artificially suppressed 
when steep density gradients are present at the contact discontinuities. 
Moreover, the level of central entropy produced in binary merger non-radiative 
simulations of galaxy clusters is significantly higher by a factor $\sim2$ in 
Eulerian simulations than in those made using SPH codes \citep{mi09}.
The origin of these discrepancies has been recognized \citep{ag07,read10} as partly 
due to the intrinsic difficulty for SPH to properly model density gradients
across fluid interfaces, which in turn implies the presence of a surface tension 
effect which   inhibits  the growth of KH instabilities.
A second problem is due to the Lagrangian nature of SPH codes, which prevents
mixing of fluid elements at the particle level and leads to entropy 
generation  \citep{wa08,mi09}. In particular, \cite{mi09} argue that the main 
explanation 
for the different levels of central entropy found in cluster simulations is due
to the different degree of entropy mixing present in the two codes.
Unlike SPH, in Eulerian codes fluid mixing occurs by definition at the cell 
level and a certain degree of overmixing is certainly present in those simulations
 made using mesh-based codes \citep{sp10b}.

Given the advantages  of SPH codes highlighted previously, it appears worth 
pursuing any improvement in the SPH method capable of overcoming its present 
limitations. 
Along this line of investigation many efforts have been made by 
a number of authors \citep{ab11,pr08,wa08,vrd10,read10,hs10,ch10,mu11}.

\cite{pr08} introduces an artificial heat conduction term  
in the SPH equations with 
the purpose of smoothing thermal energy at fluid interfaces. This artificial 
conductivity (AC)  term in turn gives a smooth entropy transition at 
contact discontinuities with the effect of enforcing pressure continuity and 
removing the artificial surface tension effect which inhibits the growth 
of KH instabilities at fluid interfaces.
Similarly, \cite{wa08} suggest that in SPH the lack of mixing at the particle level 
can be alleviated by adding a heat diffusion term  to the equations  so as 
to  mimic the effects of subgrid turbulence, thereby improving 
the amount of mixing. 

\cite{read10} present an SPH implementation in which a modified density 
estimate is adopted \citep{rt91}, together with the use of a peaked kernel 
and a much larger number 
of neighbors. The authors showed that the new scheme is capable of following 
the development of fluid instabilities  in a better way than in standard SPH.
\cite{ab11} presents an alternative derivation of the SPH force equation which
avoids the problem encountered by standard SPH in handling fluid instabilities, 
although the approach is not inherently  energy or momentum conserving and is 
prone to large integration errors when the simulation resolution is low.

The method proposed by \cite{I02} reformulates the SPH equations by introducing a
kernel convolution so as to consistently calculate density and hydrodynamic forces.
 The latter are determined using a Riemann solver  (Godunov SPH). The method
has recently been revisited by \cite{ch10} and \cite{mu11}, who showed that the code correctly 
follows the development of 
fluid instabilities in a variety of hydrodynamic tests.

A deeper modification than those presented here so far  has been introduced
by \cite{hs10}, who replaced  the traditional SPH kernel approach with an new 
density estimate based on Voronoi tesselation. The authors showed that 
the method is free of surface tension effects and therefore the growth rate of 
fluid instabilities is not adversely affected as in standard SPH.

Finally, a radically new numerical scheme has been introduced by \cite{sp10b}, with 
the purpose of retaining the advantages of both SPH and mesh-based codes.
In the new code the hydrodynamic equations are solved on a moving 
unstructured 
mesh using a Godunov method with an exact Riemann solver. The mesh is defined 
by the Voronoi tesselation of a set of discrete points and is allowed to move 
freely with the fluid.
The method is therefore adaptative in nature and thus Galilean invariant but, 
at the same time, the accuracy with which shocks and contact discontinuities 
are described is that of an Eulerian code. 
In a recent paper \cite{ba12} argue that the standard formulation of SPH fails 
to accurately resolve the development of  turbulence in the subsonic regime 
(but see \cite{pr12b} for a different viewpoint).
 The authors draw their conclusions by analyzing results from simulations of
driven subsonic turbulence made using the new moving-mesh code, named AREPO, 
and a standard SPH code.

Similar conclusions were reached in a set of companion papers \citep{sj11,vog11}, in 
which the new code was used in galaxy formation studies to demonstrate 
its superiority over standard SPH. However, the code is characterized by  
 considerable complexity which makes the use the SPH scheme still appealing and, 
more generally, it is desirable that simulation results produced with a 
specific code should be reproduced with a completely independent numerical 
scheme when complex non-linear phenomena are involved.  It appears worthwhile, 
therefore,  to investigate, along the line of previous authors, the possibility of 
constructing a numerical scheme based on the traditional SPH formulation  
which is capable of correctly describing the development of fluid instabilities 
and at the same time incorporates the effects of self-gravity when present.

 This is the aim of the present study, in which the SPH scheme is modified by 
incorporating into the equations an AC diffusion term as described by \cite{pr08}.
However, in \cite{pr08} the strength of the AC is governed by a signal velocity 
which is based on pressure discontinuities. For simulations where gravity is 
present this approach is not applicable because hydrostatic equilibrium 
requires pressure gradients.
An appropriate signal velocity for conductivity when gravity is present is then
used to construct an AC-SPH code with the purpose of treating 
the growth of fluid instabilities self-consistently. 
The viability of the approach is tested using a suite of test problems in which 
results obtained using the new code are contrasted with the corresponding ones 
produced in previous work using different schemes.
The code is very similar in form to that presented by \cite{wa08}, but here the 
energy diffusion equation is implemented in a different manner.

 The paper is organized as follows. Sect. \ref{hymeth.sec} presents 
the hydrodynamical method and introduces the AC approach. In Sect. \ref{hydro.sec}
we investigate the effectiveness of the method by presenting results from
a suite of purely hydrodynamical test problems. These are: the two-dimensional 
 Kelvin-Helmholtz instability, the Sod shock tube, the point explosion or Sedov 
blast wave test and the blob test. In Sect.  \ref{hygrav.sec} we
then discuss results from hydrodynamic tests that include self-gravity. 
Specifically, we consider the cold gas sphere or Evrard collapse test,
the Rayleigh-Taylor instability  and the cosmological integration of 
galaxy clusters.
Finally, the main results are summarized in Sect.  \ref{conc.sec}

\section{The Hydrodynamic method}
\label{hymeth.sec}
Here we present the basic features of the method, for recent reviews 
see \cite{ro09}, \cite{sp10} and \cite{pr12}. 

\subsection{Basic SPH equations}
\label{sph.sec}

In the SPH method the fluid is described  by a set of $N$ particles with 
mass $m_i$, velocity $\vec v_i$, density
$\rho_i$, and a thermodynamic variable such as the specific thermal energy 
$u_i$ or the entropic function $A_i$. The particle pressure is then defined as
$P_i=(\gamma-1) \rho_i u_i=A_i\rho_i^{\gamma}$, where $\gamma=5/3$ for a 
monoatomic gas.
The density estimate $\rho(\vec r)$ at the particle position $\vec r_i$ 
is given by 

\be
 \rho_i=\sum_j m_j W(|\vec r_{ij}|,h_i)~,
    \label{rho.eq}
\ee

where $\vec r_{ij}\equiv \vec r_i-\vec r_j$,  $W(|\vec r_{ij}|,h_i)$ is 
the interpolating kernel
that has compact support and is zero for $|\vec r_{ij}|\geq \zeta h_i$
\citep{pr12}. The kernel is normalized to the condition $\int Wd^Dr=1$.
 The sum in Eq. (\ref{rho.eq}) is over a finite number of 
particles and the smoothing length $h_i$  is a variable
that is implicitly defined by the equation 

   \begin{equation}
  f_D(\zeta h_i)^D \rho_i=N_{sph} m_i~,
    \label{hrho.eq}
   \end{equation}

where $D$ is the number of spatial dimensions, $f_D=\pi,4\pi/3$ for $D=2,3$ 
and  $N_{sph}$ is the number of neighboring particles within a radius 
$\zeta h_i$ . This equation can be rewritten by defining $h_i$ in units 
of the mean interparticle separation 

\be
h_i=\eta (m_i/\rho_i)^{1/D}~,
  \label{hzeta.eq}
\ee

so that $N^{2D}_{sph}= {\pi (\zeta \eta)^2 }$ and 
$N^{3D}_{sph}= {4 \pi (\zeta \eta)^3 }/{3}$. The smoothing length $h_i$ is 
determined by solving the non-linear equation  (\ref{hrho.eq}); note 
that $N_{sph}$ does not 
necessarily need to be an integer but can take arbitrary values if $\eta$ is
used as the fundamental parameter determining $h_i$ \citep{pr12}.  
A kernel commonly employed is the $M_4$, or cubic spline, which is zero 
for $\zeta\geq2$. In three dimensions typical choices for $N_{sph}$ 
lie in the range $N_{sph}\sim 33-64$, which for the $M_4$ kernel
corresponds to $\eta\simeq 1.-1.25$.
The equation of motion for the SPH particles can be derived from the 
Lagrangian of the system \citep{sp10,pr12} and is given by 

 \be
  \frac {d \vec v_i}{dt}=-\sum_j m_j \left[
  \frac{P_i}{\Omega_i \rho_i^2}
  \vec \nabla_i W_{ij}(h_i) +\frac{P_j}{\Omega_j \rho_j^2}
   \vec \nabla_i W_{ij}(h_j)
   \right]~,
  \label{fsph.eq}
  \ee
where the coefficients $\Omega_i$ are defined as 

 \be
 \Omega_i=\left[1-\frac{\partial h_i}{\partial \rho_i}
 \sum_k m_k \frac{\partial W_{ik}(h_i)}{\partial h_i}\right].
  \label{fh.eq}
 \ee

These terms are present in the  momentum equation (\ref{fsph.eq})
because the smoothing length $h_i$ itself is implicitly a function of 
the particle coordinates through Eq. (\ref{hrho.eq}).

\subsection{Artificial viscosity}
\label{avvis.sec}

In SPH, the momentum equation must be 
generalized by adding an appropriate viscous force, which  is aimed
at correctly treating the effects of shocks. 
An artificial viscosity (AV) term is then introduced with the purpose 
of dissipating local velocities  and preventing particle interpenetration at
the shock locations. The new term is given by

 \be
\left (\frac {d \vec v_i}{dt}\right )_{AV}=-\sum_i m_j \Pi_{ij} \vec \nabla
_i \bar W_{ij}~,
    \label{fvis.eq}
  \ee

where the term  $\bar W_{ij}= \frac{1}{2}(W(r_{ij},h_i)+W(r_{ij},h_j))$ is the
symmetrized kernel and $\Pi_{ij}$ is the AV tensor.
In the SPH entropy formulation \citep{sh02}, it is the entropy  function per 
particle $A_i$ which is integrated and its time  derivative is calculated 
as follows
 
   \begin{equation}
  \frac {d A_i}{dt} =\frac{1}{2}\frac{\gamma-1}{\rho_i^{\gamma-1}}
  \sum_j m_j \Pi_{ij} \vec v_{ij}\cdot \nabla_i \bar W_{ij}\equiv
  \frac{\gamma-1}{\rho_i^{\gamma-1}} 
\left (\frac{d u_i}{dt} \right )_{AV}~,
    \label{avis.eq}
   \end{equation}

 where $\vec v_{ij}= \vec v_i - \vec v_j$. For the AV tensor we adopt here the 
form proposed by \citep{mo97} in analogy with Riemann solvers:

 \be
\Pi_{ij} = 
 -\frac{\alpha_{ij}}{2} \frac{v^{AV}_{ij} \mu_{ij}} {\rho_{ij}} f_{ij}~,
  \label{pvis.eq}
\ee

where 
\be
v^{AV}_{ij}= c_i +c_j - 3 \mu_{ij}
  \label{vsig.eq}
 \ee
is the signal velocity 
and  $\mu_{ij}= \vec v_{ij} \cdot \vec r_{ij}/|r_{ij}|$ if 
$ \vec v_{ij} \cdot \vec r_{ij}<0$ but zero otherwise,
 so that  $\Pi_{ij}$ is non-zero only for approaching particles. 
  Here scalar quantities with the subscripts $i$ and $j$ denote arithmetic 
averages, $c_i$ is the sound speed of particle $i$, the parameter $\alpha_i$
  regulates the amount of AV, and $f_i$ is a controlling factor that 
reduces the strength of AV in the presence of shear flows. 
The latter is given by \citep{ba95}

   \be
  f_i=\frac {|\vec \nabla \cdot \vec v|_i}
  {|\vec \nabla \cdot \vec v|_i+|\vec \nabla \times \vec v|_i}~,
   \label{fdamp.eq}
   \ee

where $(\vec \nabla \cdot \vec v)_i$ and $(\vec \nabla \times \vec v)_i$
are the standard SPH estimates for divergence and curl \citep{mo05}.
For pure shear flows 
$|\vec \nabla \times \vec v|_i>> |\vec \nabla \cdot \vec v|_i$ 
so that the AV is strongly damped.

Using the signal velocity (\ref{vsig.eq}), the Courant condition on the 
timestep  of particle $i$ reads
\be
\Delta t^{C}_i \simeq 0.3 \frac{h_i} {max_j |v^{AV}_{ij}|}.
 \label{cour.eq}
 \ee

In the standard SPH formulation  the viscosity  parameter $\alpha_i$ which 
controls the strength of the AV is given by 
$\alpha_i=\mathrm{const}\equiv \alpha_0$,
with $\alpha_0=1$ being a common choice  \citep{mo05}.
This scheme has the disadvantage that it generates viscous dissipation in 
regions of the flow that are not undergoing shocks. 
In order to reduce spurious viscosity effects \cite{mm97} proposed 
  making the viscous coefficient $\alpha_i$  time dependent so that it can 
 evolve in time under certain local conditions. The authors proposed an 
equation of the form

 \be
  \frac {d \alpha_i}{dt} =-\frac{\alpha_i-\alpha_{min}}{\tau_i} +{S}_i~,
    \label{alfa.eq}
 \ee

 where ${S}_i$ is a source term  and $\tau_i$ regulates the decay
of  $\alpha_i$ to the floor value $\alpha_{min}$ away from shocks.
For the source term the following expression is adopted 
   \begin{equation}
 {S}_i=S_0 f_i (-(\vec \nabla \cdot \vec v)_i,0)
(\alpha_{max}-\alpha_i)~,
    \label{salfa.eq}
   \end{equation}

 which is constructed in such a way that it increases in the presence of 
 shocks. The prefactor $S_0$ is unity for $\gamma=5/3$ and the damping factor
 $f_i$ is inserted to account for the presence of vorticity.  
The original form proposed by \cite{mm97} has been refined by the 
factor $(\alpha_{max}-\alpha_i)$ \citep{ro00}, which has the advantage of 
being more  sensitive to shocks than the original formulation and of preventing 
$\alpha_i$ from becoming higher than a prescribed value $\alpha_{max}$.
Recently, \cite{cul11} presented an improved version of the \cite{mm97} AV scheme,
employing as shock indicator a switch based on the time derivative of the 
velocity divergence.

The decay parameter $\tau_i$ is of the form

   \begin{equation}
  \tau_i=\frac{h_i}{c_i ~l_d}~,
    \label{tau.eq}
   \end{equation}

where $l_d$ is a dimensionless parameter which controls the decay
time scale.
In a number of test simulations, \citep{ro00} found that appropriate values 
for the parameters $\alpha_{max},\alpha_{min}$, and $l_d$ are $1.5,0.05$, and 
$0.2$, respectively.  In principle, the effects of numerical viscosity in
regions away from shocks can be reduced by setting $l_d$ to higher values than 
 $l_d=0.2$. In practice, the minimum time necessary to propagate through 
the resolution length $h_i$ sets the upper limit $l_d=1$.
However, the time evolution of the viscosity parameter
can be affected if  very short damping timescales are imposed.
Neglecting variations in the coefficients, the solution to Eq.
(\ref{alfa.eq}) at times $t> t_{in}$ can be written as

   \be
  \alpha_i(t)=q_i+(\alpha_i(t_{in}) -q_i)\exp ^{-(t-t_{in})/\tau^{\prime}_i}~,
   \label{alfatn.eq}
   \ee

where
   \be
  \tau^{\prime}_i=\frac{\tau_i}{1+S_i \tau_i}~,
    \label{taunew.eq}
   \ee

and  $q_i$ is a modified source term

   \be
 q_i=\frac{\alpha_{min}+S_i\tau_i \alpha_{max}}{1+S_i\tau_i}.
    \label{qsou.eq}
   \ee

From Eq. (\ref{alfatn.eq}) it can be seen that 
$\alpha_i(t)  \simeq \alpha_{max}$ in the strong shock regime 
$S_i \tau_i \gg1$ but this condition is not satisfied if $S_i \tau_i \simlt 1$.
Therefore for mild shocks this implies that, 
if very short decay time scales are imposed,
the peak value of $\alpha_i(t)$ at the shock front might be below the AV 
strength necessary to properly treat shocks.

To overcome this difficulty a modification to (\ref{salfa.eq}) has been 
adopted (Valdarnini 2011, herefater V11) which, when $l_d=1$, compensates 
 the smaller values of 
$S_i\tau_i$ with respect to the small $l_d$ regimes.
 This is equivalent to considering a higher value for $\alpha_{max}$, so that 
in Eq. (\ref{salfa.eq}) $\alpha_{max}$ is substituted by 
$\alpha_{max}\rightarrow \xi \alpha_{max}$.

\begin{table}
%\begin{center}
\caption{KH parameters for the simulations. From top to bottom: 
simulation label, Mach number $M$ and $x-$velocity $v_1$ of the 
high-density layer, KH time scale $\tau_{KH}$}
\def\arraystretch{1.1}
\begin{tabular}{lrrrrr}
\hline
%Scheme & Time per step $\left[\s\right]$ & Total steps & Total CPU hours & CFL\\
\hline
\textsf{label run}    &  1  &  2  &  3  &  4 & 5 \\
\hline
\textsf{M}  &  0.2  &   0.4  &   0.6  &  0.8 & 1.0 \\
$|v_x|$  &  0.26  &   0.52  &   0.77  &  1.0  & 1.3 \\
$\tau_{KH}$  & 1.23   &  0.56   &  0.37   & 0.28  & 0.22 \\
%\textsf{HLL5R}  &  $67.9\pm0.5$  &   2,065  &   2,493  &  0.8 \\
\hline
\\
\end{tabular}
\label{tab:mach}
%\end{center}
\end{table}

The correction factor $\xi$ has been calibrated  using  
 the shock tube problem as reference and requiring a peak value of 
$\sim0.6-0.7$ for the viscosity parameter at the shock front, as in the 
$l_d=0.2$ case. The results indicate that a normalization of the form 
$\xi=(l_d/0.2)^{0.8}$ for $ l_d\geq0.2$ satisfies these constraints.
 This normalization  has been validated in a number
of other test problems showing that it is able to  produce
 a  peak value of the viscosity parameter at the shock location
  which is independent of the chosen value of the decay parameter $l_d$.

The time-dependent AV formulation of SPH has been shown to be effective 
in reducing the  damping of turbulence due to the effects of 
numerical viscosity in simulations of galaxy clusters  \citep[V11]{dl05}
%(Dolag et al. 2005; V11).
Moreover, it has been used in a recent paper \citep{pr12b} to argue that the 
conclusion of \cite{ba12} about the difficulty for SPH codes in properly 
modeling the development of subsonic turbulence is based just on  
 using a SPH code in its standard AV formulation.
In the following, unless otherwise stated,  simulations will be performed 
 using a time-dependent AV; this will be fully specified by the  set of 
parameters $\{ \alpha_{min}, \alpha_{max}, l_d\}$.

\subsection{Artificial conductivity}
\label{acsph.sec}
As already outlined in the Introduction, different formulations have been 
proposed for overcoming the problems encountered by standard SPH in the 
treatment of fluid discontinuities. Here we follow the approach suggested
by \cite{pr08}, who proposed adding a dissipative term to the thermal energy
equation for smoothing the energy across contact discontinuities.
The presence of these dissipative terms introduces a smoothing of entropy
at fluid interfaces  
with the effect of removing pressure discontinuities.

The motivations lying at the basis of this approach were discussed 
by \cite{mo97}, who showed how the momentum and energy equations in SPH 
must contain a dissipative term 
related to jumps in the variables across characteristics, in analogy with the 
corresponding Riemann solutions. 

The artificial conductivity (AC) term for the dissipation of energy takes the 
form

 \be
  \left ( \frac {d u_i}{dt} \right)_{AC} = 
\sum_j \frac{m_j v^{AC}_{ij}}{\rho_{ij}} 
\left[ \alpha^C_{ij}(u_i-u_j) \right ] \vec {e_{ij}}\cdot \vec {\nabla_i}
 \bar W_{ij}~,
  \label{duc.eq}
 \ee

where $ v^{AC}_{ij}$ is the signal velocity, which does not need to be the same 
as that used in the momentum equation, 
$\vec e_{ij} \equiv \vec r_{ij}/r_{ij}$
and $\alpha^C_{i}$ is the AC parameter
which is of  order unity \footnote{Note that there is a misprint in the sign of 
the corresponding Eq. 28 of \cite{pr08}}.
 Eq. (\ref{duc.eq}) represents the SPH analogue of a diffusion 
equation of the form 

 \be
  \left ( \frac {d u_i}{dt} \right)_{AC}  \simeq D^{AC}_i \nabla^2 u_i~,
  \label{dudis.eq}
 \ee

where 

\be
\nabla^2 u_i=2\sum_j m_j \frac{u_i-u_j}{\rho_j}\frac{\vec e_{ij}\cdot
\vec {\nabla} W_{ij}}{r_{ij}}
 \label{udii.eq}
\ee

is the SPH expression for the Laplacian \citep{br85} and, in analogy with 
the analysis of \cite{lp10} for defining an equivalent physical viscosity 
coefficient for the SPH numerical viscosity, we 
can define $D^{AC}_i$ as a numerical heat diffusion coefficient
given by

\be
 D^{AC}_i \simeq \frac{1}{2}\alpha^C_{i} v^{AC}_{ij} r_{ij}~.
 \label{dac.eq}
\ee

\begin{table}
\caption{ Main simulation parameters of the KH tests. From top to bottom: 
simulation label, 
kernel function, neighbor number, signal velocity used in Eq. (\ref{duc.eq}), 
setup of initial HCP lattice: U=unsmoothed, S=smoothed. For all of the runs
the AV parameters are $\{ \alpha_{min}, \alpha_{max}, l_d\}=\{0.1,1.5,1\}$.
}
\def\arraystretch{1.1}
\begin{tabular}{lrrrrrr}
\hline
\hline
\textsf{Simulations}    &  \textsf{SPH}  &  \textsf{RHO}  & \textsf{LIQ}  
  & \textsf{LP} & \textsf{CRT}  &  \textsf{M5} \\
\hline
\textsf{Kernel}  &   \textsf{CS}  &  \textsf{CS}    &   \textsf{LIQ}  &
\textsf{LIQ}  & \textsf{CRT}  &  \textsf{M5} \\
$N_s$  &  32  &   32  &   32  &  32  & 32  & 50 \\
\textsf{$v^{sig}$}   & \textsf{grav}   & \textsf{grav}     &  \textsf{grav}    & 
\textsf{pres} & \textsf{grav}   & \textsf{grav}  \\
\textsf{$\rho$}   & \textsf{U}   & \textsf{S}     &  \textsf{S}    & 
\textsf{S} & \textsf{S}   & \textsf{S}  \\
\hline
\\
\end{tabular}
\label{tab:cases}
\end{table}
An important issue concerns the choice of the AC parameter $\alpha^C_{i}$ which 
must  be constructed so that diffusion of thermal energy is kept under 
control and is damped away from discontinuities. 
This can be achieved by introducing a switch 
similar to that devised for AV. The dissipation parameter is then evolved 
according to 

 \be
\frac {d \alpha^C_i}{dt} =-\frac{\alpha^C_i-\alpha^C_{min}}{\tau^C_i} +{S^C}_i~,
    \label{alfac.eq}
 \ee
the meaning of the terms being similar to that of Eq. (\ref{alfa.eq}). In 
particular, for the decay timescale $\tau^C_i={h_i}/{\mathcal{C} c_i }$
we set here $\mathcal{C}=0.2$ and set to zero the floor value $\alpha^C_{min}$.
For the source term $S^C_i$ the following expression is used

\be
{S^C}_i=f_C h_i\frac{|\nabla^2 u_i|}{\sqrt{u_i+\varepsilon}} 
(\alpha^C_{max}-\alpha^C_i)~,
    \label{salfac.eq}
\ee

where $f_C$ is a dimensionless parameter of order unity and the parameter 
$\varepsilon$ avoids divergences as $u_i\rightarrow 0$. This equation differs 
from the corresponding  source term proposed by \cite{pr05}, \cite{pr08} by 
the factor
$\alpha^C_{max}-\alpha^C_i$ which has been inserted here in analogy with Eq.
(\ref{salfa.eq}) for introducing a stronger response 
of the switch in the presence of discontinuities. The choice of values for
the parameters  $f_C$  and $\alpha^C_{max}$ depends on the problem under 
consideration, 
for example \cite{pr05} proposed $f_C=0.1$. Here a number of numerical 
experiments has shown that the best results in terms of response of the 
AC parameter $\alpha^C_i$ to the presence of discontinuities are 
obtained by setting $f_C=1$  and $\alpha^C_{max}=1.5$, which we assume 
henceforth as reference values. Moreover it is found that significant benefits 
in terms of sharpness of the AC parameter profile across the discontinuity 
are obtained by inserting the $\alpha^C_{max}-\alpha^C_i$ term.
In principle, the choice of the derivative term used to detect discontinuities 
is arbitrary, but in practice  a second derivative \citep{pr05} term 
ensures better sensitivity to sharp discontinuities in thermal energy than a 
first derivative \citep{pr05a}.

An example of a signal velocity specifically designed to remove pressure 
gradients at 
contact discontinuities is that originally introduced by \cite{pr08} 
for pure hydrodynamical simulations
\be
v^{AC}_{ij}(P) = \sqrt{\frac{|P_i-P_j|}{\rho_{ij}}}~,
  \label{vspr.eq}
 \ee
and further refined by \cite{vrd10}. The ability of the new AC formulation 
of SPH to follow the development of KH instabilities using this expression
for the signal velocity has been  verified in a number of tests 
\citep{pr08,vrd10,me10}.  However, the disadvantage of the signal speed 
(\ref{vspr.eq}) is that it cannot be applied when self-gravity is considered 
because in that case a pressure gradient is present at 
hydrostatic equilibrium. {Imagine, for example, a self-gravitating 
gas sphere 
in hydrostatic equilibrium in which there is  a negative radial temperature 
gradient, with the gas  temperature decreasing  when moving outward from the
 center of the sphere.  An application of the SPH equations, with the AC term  of
Eq. (\ref{alfac.eq}) now using the signal velocity (\ref{vspr.eq}), will lead 
to a heat flux from the inner to the outer 
regions and, in the long term, to the development of an unphysical isothermal
temperature profile. }
A signal velocity which avoids this difficulty is given by

\be
v^{AC}_{ij}(grav) = |(\vec v_i-\vec v_j)\cdot \vec r_{ij}|/r_{ij}~,
 \label{vsgv.eq}
 \ee
which corresponds to the formulation proposed by \cite{wa08} in which a 
dissipative term is added to the evolution of the thermal equation with
the purpose of modeling heat diffusion due to turbulence. The new term is 
constructed in analogy with the subgrid-scale model of \cite{sm63} and
is given by  
 
 \be
 \left( \frac {d u_i}{dt} \right)_{AC}= C \sum_j \frac{ m_j|\vec v_i-\vec v_j|(h_i+h_j) }
{\rho_{ij}} (u_i-u_j)  \vec e_{ij}\cdot \vec {\nabla_i}
 \bar W_{ij}~,
  \label{ducw.eq}
 \ee
where $C$ is a coefficient of order unity, whose precise value depends on the 
problem under consideration. For the rising hot bubble problem considered 
by \cite{wa08}, the best agreement with the corresponding PPM results taken 
as reference is 
obtained setting $C=0.1$, higher values being too much diffusive. Throughout this
paper the SPH simulations of the hydrodynamic tests are performed by adopting 
the expression (\ref{duc.eq}) for the thermal energy dissipative term. With respect
to the formulation of \cite{wa08} this approach presents the advantage of using a 
diffusion parameter which is not constant but evolves in time according to a
source term, so that the amount of diffusion away from discontinuities is 
minimized. For the AC signal velocity, we then use expression (\ref{vsgv.eq}), 
whose performances in the AC formulation (\ref{duc.eq}) has not  
been fully tested before in SPH simulations of hydrodynamic test problems.
 
Note that, using the signal velocity (\ref{vsgv.eq}), the Von Neumann stability
criterion becomes unimportant with respect  the Courant condition 
(\ref{cour.eq}).
The Von Neumann constraint requires $\Delta t \leq 0.5 \Delta x^2 /D$, which in
SPH reads 
\be
\Delta t_i^{AC}\simlt \frac{1}{2} \frac{h^2_i}{D_i^{AC}}~.
 \label{dtac.eq}
 \ee
Since $\alpha^C\simeq O(1)$, this condition implies $\Delta t_i^{AC}\simgt 
\Delta t_i^C$
 in both the supersonic and subsonic regimes.
\begin{figure*}
\centerline{
\includegraphics[width=17.2cm,height=15.2cm]{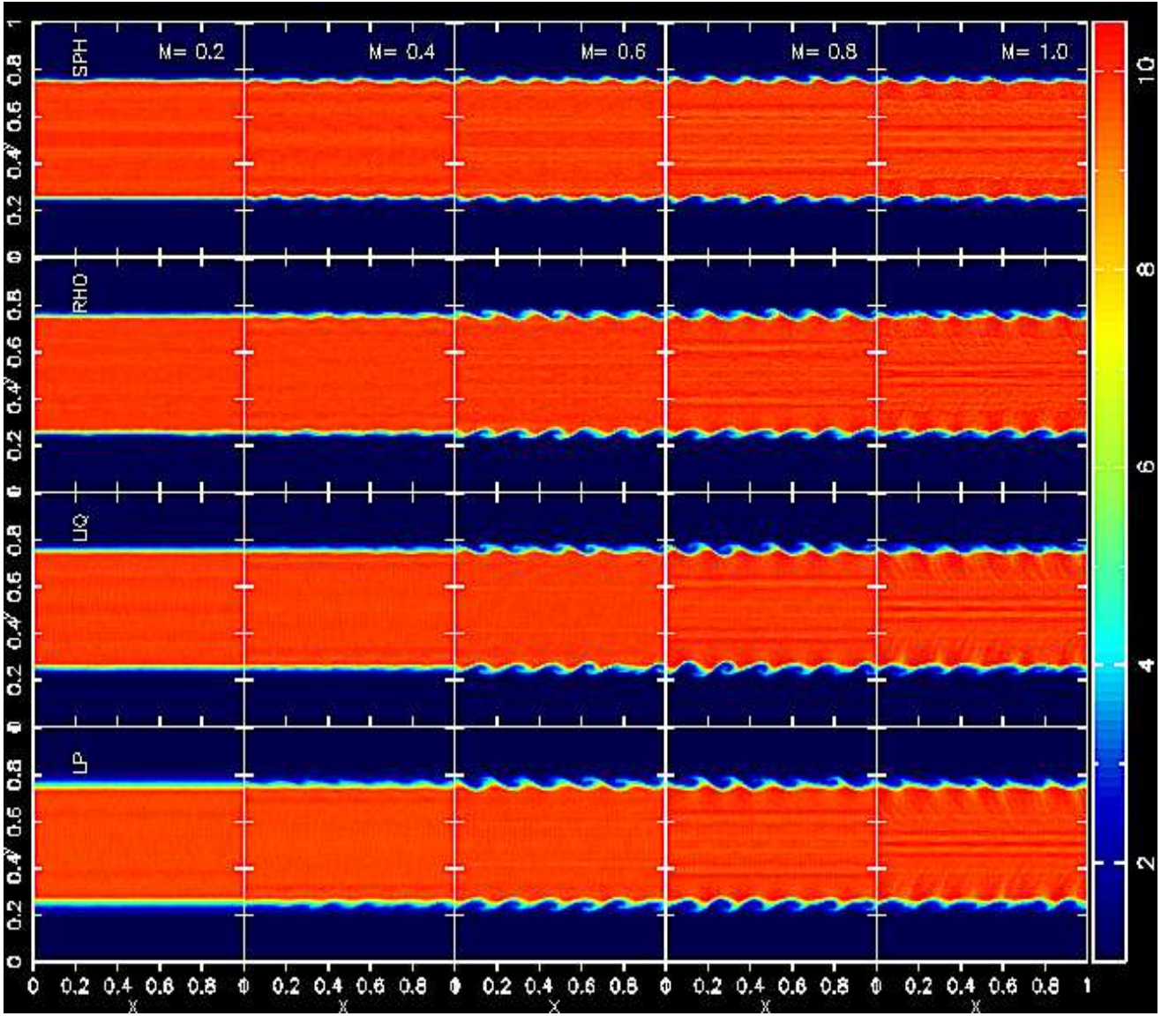}
}
\caption{Density maps for some of the 2D KH instability tests described in Sect. 
\ref{kh2d.sec}. From left to right, each column shows the panels for runs having 
initial conditions with the same Mach number at the corresponding time 
$t=\tau_{KH}$, as listed in Table \ref{tab:mach}.
 From top to bottom, different rows refer to the following
simulations : SPH, RHO, LIQ and LP. The latter use the linear quartic kernel 
but with the signal velocity (\ref{vspr.eq}), whereas the first three  use
the expression (\ref{vsgv.eq}) (see Table \ref{tab:cases}). 
The plots can be compared directly with Fig. 10 of \cite{vrd10}.}
\label{fig:vrdmap_1}
\end{figure*}

\section{ Hydrodynamic tests}
\label{hydro.sec}
In the following, simulation results obtained by applying the new AC-SPH code
to a number of hydrodynamic test problems are discussed  with the objective
of validating the code and assessing its performance. To this end, the simulation
results of the tests will be compared with the corresponding ones obtained 
by previous authors using different codes and/or numerical methods.
The problems considered are usually presented in the literature in dimensional 
or complexity order, but here we follow a different approach. 
Since the KH instability is the  hydrodynamic test which initially 
\citep{ag07} originated the discussion about the difficulties of standard SPH 
in properly handling the development of instabilities, and 
 it is also the most widely considered in this context  
\citep{ab11,pr08,ch10,hs10,read10,vrd10,mu11}, we here  discuss first in detail the 
two-dimensional version of the test.  The setup of the other runs will then 
be considered in the light of the results obtained from the 2D KH test.

\subsection{2D Kelvin-Helmholtz instability}
\label{kh2d.sec}
 The KH instability arises in the presence of shear flow between two fluid 
layers when a small velocity perturbation is imposed in the direction 
perpendicular to the interface between the two fluids.
The development of the instability is characterized by an initial phase, where 
the fluids interpenetrate each other, and then the forming of vortices, which 
become progressively more pronounced and turn into KH rolls  at 
the onset of non-linearity.
In the case of incompressible fluids for  a sinusoidal perturbation of 
wavelength $\lambda$, this phase is reached
with a time-scale \citep{ch61}

\be
\tau_{KH}=\frac{\lambda (\rho_1+\rho_2)}
{\left(\rho_1\rho_2\right)^{1/2}v}~,
\label{taukh.eq}
\ee

where $\rho_1$ and $\rho_2$ are the fluid densities and $v=v_1-v_2$ is the 
relative shear velocity.
A proper modeling of instabilities in numerical hydrodynamics is essential
since the KH instability plays a crucial role in the development of 
turbulence which occurs in many hydrodynamical phenomena.  In particular, 
the KH instability  is relevant in many astrophysical processes, 
such as gas stripping from cold gas clouds occurring in galactic haloes 
and the production of entropy in the intracluster gas of galaxy 
clusters due to injection of turbulence.

\begin{figure*}
\includegraphics[width=17.2cm,height=11.5cm]{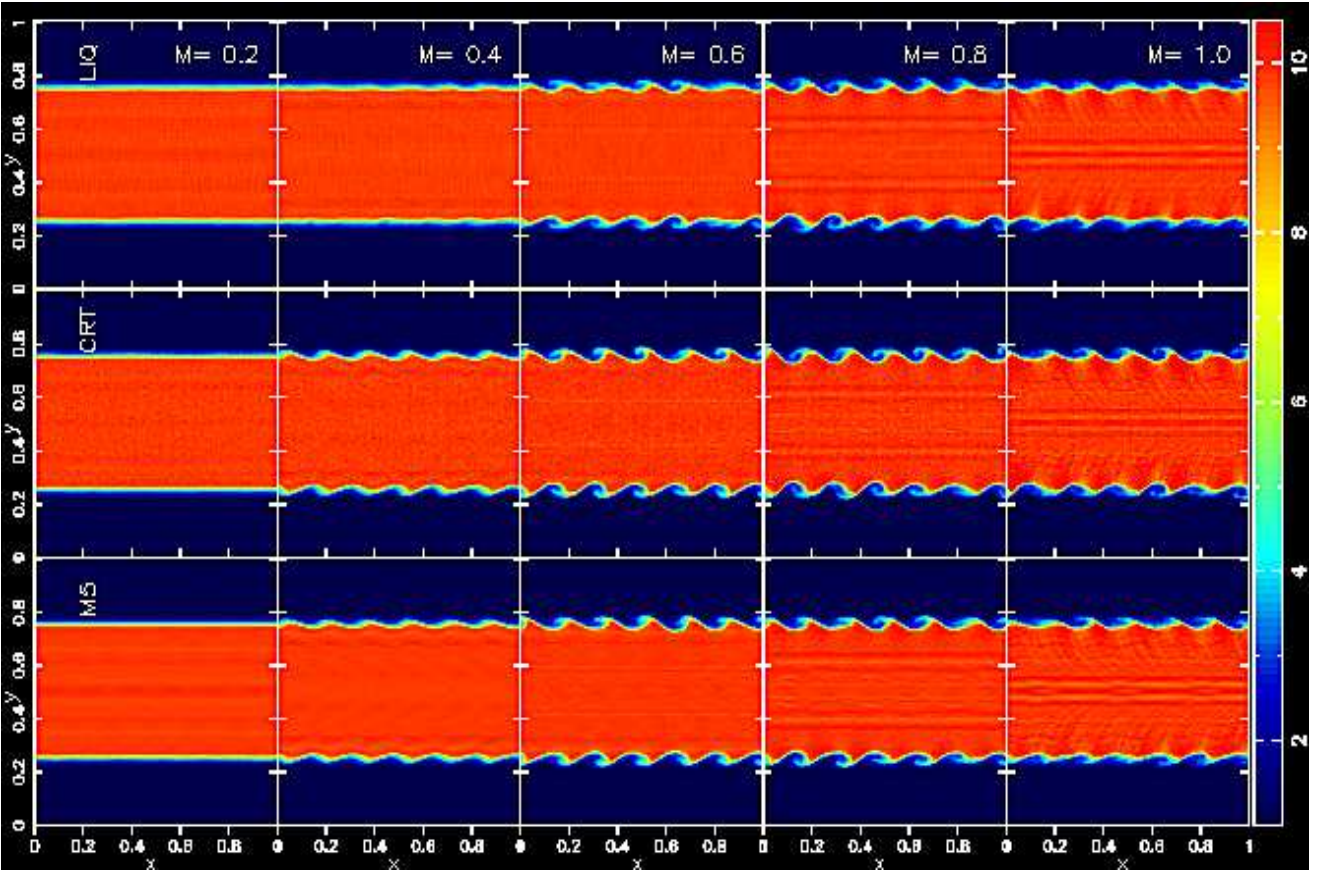}
\caption{As in Fig. \ref{fig:vrdmap_1} but for the set of simulations
LIQ, CRT and $M_5$.}
\label{fig:vrdmap_2}
\end{figure*}
The  growth of the KH instability in hydrodynamic simulations has been  addressed
by  various authors \citep{ab11,pr08,wa08,read10,vrd10,hs10,ch10,mu11}. 
These studies found that the development of the instability is 
artificially suppressed because of two
distinct effects: the first problem is the so-called local mixing instability (LMI)
 which is due to entropy conservation and which inhibits mixing at the kernel scale 
thus introducing pressure discontinuities; the second problem arises because 
of errors in the momentum equation which cannot be reduced by increasing 
the number of neighbor particles without causing particle clumping.

Given the wide variety of numerical parameters and initial conditions with 
which
the KH instability has been addressed, we choose here to perform the tests using as 
reference the simulations presented by \cite{vrd10}. 
In particular, the authors 
implement in their SPH equations an AC term as that of Eq.~(\ref{duc.eq}), 
but with the AC parameter set to unity, and with a signal velocity given by 
Eq.~(\ref{vspr.eq}).
The authors performed
a systematic analysis of the capabilities of SPH to capture the KH instability 
using different SPH formulations, kernels, numerical resolutions and KH time-scales.
This choice allows to assess code capabilities by constructing a large 
 suite of simulations whose  numerical results can be contrasted 
against the corresponding ones discussed by \cite{vrd10}.

\subsubsection{Initial conditions set-up}
\label{inkh2d.sec}
The problem domain consists of a periodic box of unit length with 
cartesian coordinates $x\in \{ 0,1 \}$, $y\in \{ 0,1 \}$. Within the 
domain  there is a fluid with adiabatic index $\gamma=5/3$ which 
satisfies the following  conditions:  

\be
\rho,~T,v_x=\left\{
 \begin{array}{ l l }
\rho_1,T_1,v_1  & |y-0.5|\leq 0.25 \\
\rho_2,T_2,v_2  & |y-0.5| >  0.25.
 \end{array}
\right.
\label{rhokh.eq}
\ee

As in \cite{vrd10},  we choose here  $\rho_1=10$ and $\rho_2=1$, with the index $1$
referring to the high density layer. This choice is motivated by the finding
that the difficulties of SPH in reproducing KH instabilities increase
as the density contrast between the two fluid layers gets higher.
The two layers are in pressure equilibrium with $P_1=P_2=10$ 
so that the sound velocities in the two layers are 
  $c_1=\sqrt{{\gamma P_1}/{\rho_1}}=1.29$ and 
 $c_2=\sqrt{{\gamma P_2}/{\rho_2}}=4.08$ , respectively.
The  layers  slide against each other with opposite shearing 
velocities $v_1=-v_2$ and 
in order to seed the KH instability a small single-mode velocity perturbation 
 is imposed along the $y-$direction  

\be
v_{y}=w_0 \sin(2\pi x/\lambda)~,
\label{eq:vy}
\ee

where   $w_0=0.025$  and $\lambda=1/6$. In order to restrict the perturbation 
to  spatial regions in the proximity of the interfaces, the perturbation 
(\ref{eq:vy}) is applied only if $|y-\sigma|<0.025$, where $\sigma$ takes the 
values of $0.25$ and $0.75$, respectively.
With this choice of parameters, the Mach number of the  high-density layer 
is $M\simeq v_1/c_1 \simeq v_1/1.29 $ and the KH time-scale is
$\tau_{KH}\simeq 0.29/v_1$. As in \cite{vrd10}, we consider KH simulations with
a range of five different Mach numbers; Table \ref{tab:mach} reports the values 
of $M$ together with the respective values of $v_1$ and $\tau_{KH}$.

In order to implement the the density set-up (\ref{rhokh.eq}), a two-dimensional 
 lattice of equal mass particles is placed inside the simulation box. 
We adopt here an isotropic hexagonal-close-packed (HCP) configuration for the 
particles coordinates instead  of the more commonly employed Cartesian grid. 
The advantage of this configuration is that, for a given number of neighbors, 
 it gives a better density estimate due to its symmetry properties.
To construct  the initial density configuration (\ref{rhokh.eq}),   
the lattice spacing of the particles is varied until the SPH density estimate
(\ref{rho.eq}) satisfies the required values within a certain tolerance 
criterion ($\simlt 1\%$) for the relative density error.
The  simulations were run using a total number of $N=512^2$ particles and
the initial particle specific energies were assigned after the density 
calculation so as to satisfy pressure equilibrium.
This particle number is larger than that used  in the runs of \cite{vrd10} 
($\simeq 200K$), but guarantees a density setup with the specified tolerance 
criterion.
\begin{figure*}
\centerline{
\includegraphics[width=17.2cm,height=11.5cm]{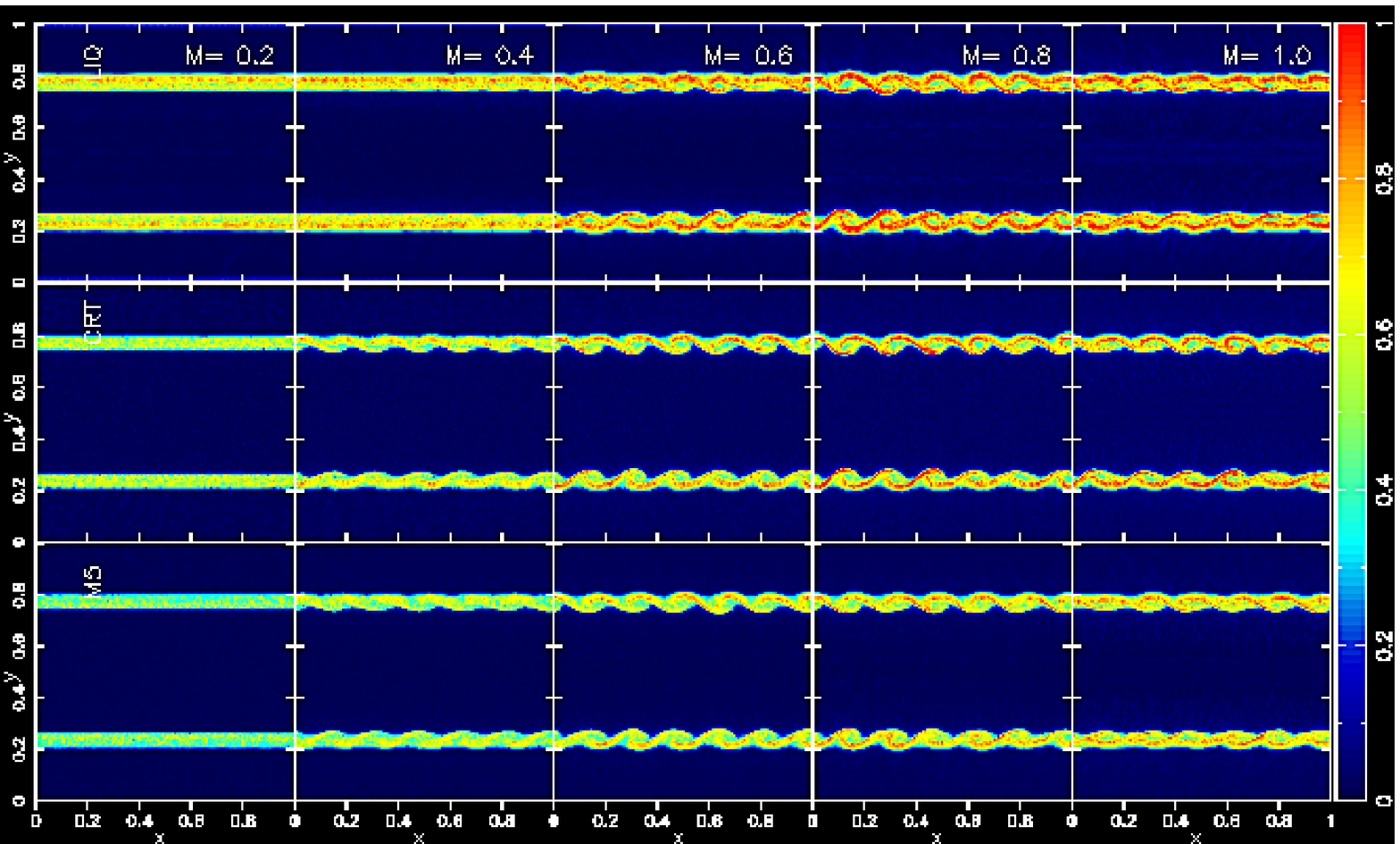}
}
\caption{Rendered plots of the $\alpha^C$ parameters are shown for the
same runs as in  Fig. \ref{fig:vrdmap_2}. The distribution  
of particle values   $\alpha^C_i$ has been interpolated at the map
grid points according to the SPH prescription.}
\label{fig:acmap}
\end{figure*}

As noticed by \cite{vrd10} and other authors, SPH is a numerical method which 
can only represent smoothed quantities, and so applying it to hydrodynamic 
problems where strong density gradients are present can lead to inconsistencies.
This is, in fact, the situation encountered by standard SPH in the treatment of KH 
instabilities, where the lack of entropy mixing induces 
an artificial pressure discontinuity  at fluid interfaces with a jump in 
density.

Motivated by these difficulties, we consider here a set of runs in which the 
density
discontinuity at the interfaces is replaced by a smooth transition. 
To allow making a consistent comparison with the corresponding runs 
of \cite{vrd10}, we adopt the same smoothing profile

\be
\rho(y)=D\pm A  \mathrm{atan} \left[ B(y^{\prime}+C) \right]~,
\label{khsm1.eq}
\ee

where the coefficients are given by

\be
A = \frac{\rho_1-\rho_2}{2 \mathrm{atan}(\beta)}~,
\label{khsm2.eq}
\ee

\be
B = 2 \frac{\beta}{\delta}~,
\label{khsm3.eq}
\ee

\be
C = - \frac{\delta}{2}~,
\label{khsm4.eq}
\ee

\be
D = \frac{\rho_1+\rho_2}{2}~,
\label{khsm5.eq}
\ee

and $y^{\prime}=y-\sigma+\delta/2$. In Eq. (\ref{khsm1.eq}) the sign in front
of the coefficient $A$ refers to $\sigma=0.25,~0.75$ respectively. The parameters 
$\beta$ and $\delta$  determine the thickness of the density transition 
and take the values $\beta=10$, $\delta=0.5/0.75$.

\begin{figure*}
\centerline{
\includegraphics[width=17.2cm,height=15.2cm]{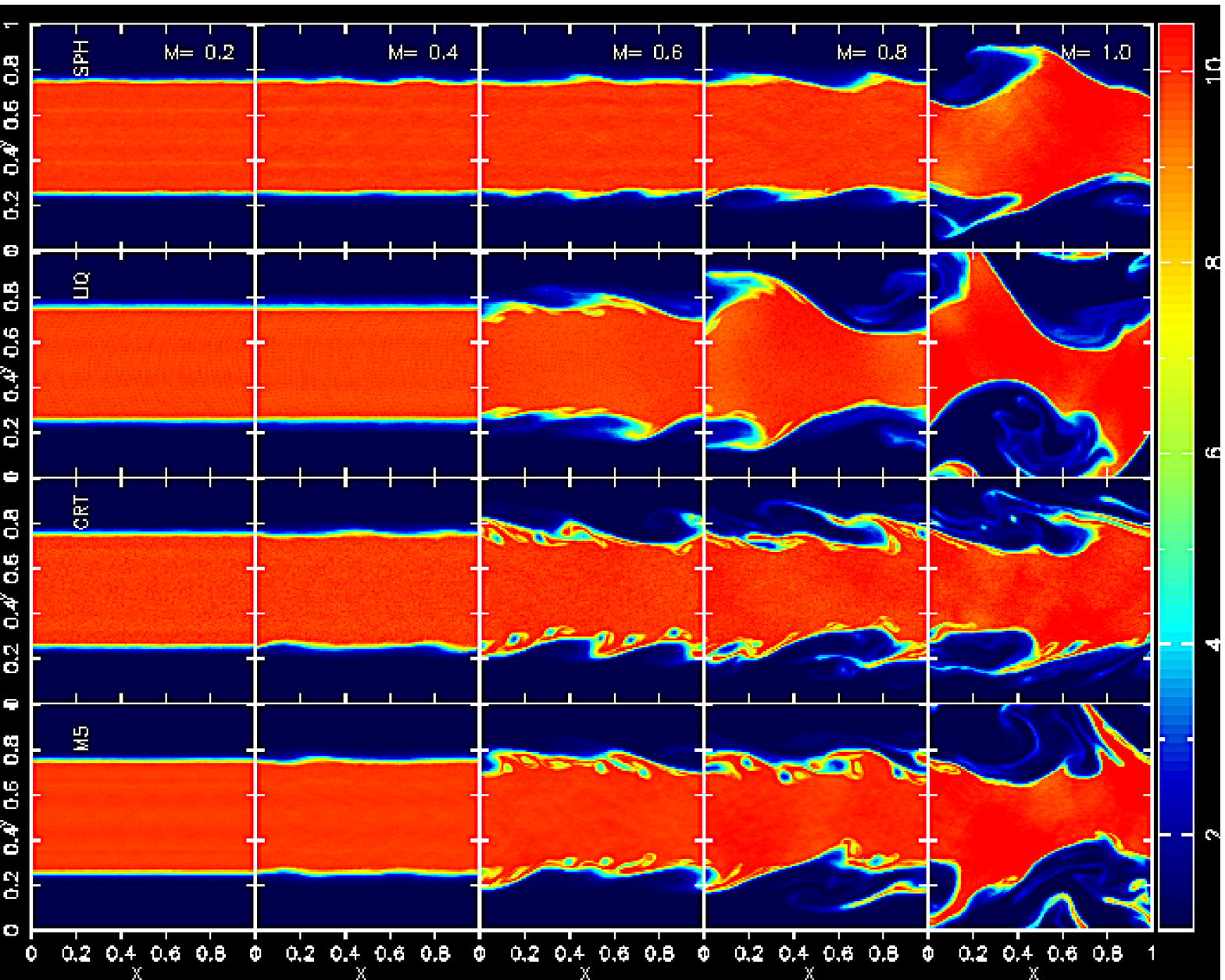}
}
\caption{As in Figures~\ref{fig:vrdmap_1} and \ref{fig:vrdmap_2} 
but here at the time $t=2$ for all of the panels. Note from Table \ref{tab:mach} 
that this implies $t>>\tau_{KH}$ for runs with high Mach number.}
\label{fig:vrdmap_3}
\end{figure*}

The following procedure is adopted in order to construct a lattice configuration 
which satisfies the density profile given by Eq. (\ref{khsm1.eq}).
An HCP lattice of particle is first constructed in the domain $0\leq x<1$ and 
$0\leq y \leq0.5$ with a spacing $a_{down}$ such that the SPH density 
is $\rho=\rho_2$.  Proceeding upward from $y=0$, the lattice spacing is 
progressively reduced according to $a=a_{down} [\rho_2/\rho(y)]^{1/2}$ 
until $y=0.25$ and $N_{down}$ particles are left. 
The same procedure is applied to a high-density lattice  which  in the 
same domain satisfies the condition $\rho=\rho_1$: starting from $y=0.5$ and 
proceeding  downward, the lattice spacing is increased so that 
$a=a_{up} [\rho_1/\rho(y)]^{1/2}$ until $y=0.25$  and $N_{up}$ particles
remain of the original high-density lattice. 
The whole procedure is numerically iterated until the numbers $N_{down}$ 
and $N_{up}$ satisfy the conditions $N_{down}+N_{up}=N/2$  and 
$N_{up}/N_{down}=\rho_1/\rho_2$.
The lattice is then replicated in the top half of the box ($y\rightarrow1-y$)
so that the initial 
conditions are fully symmetric around $y=0.5$.
In the following, simulations with initial conditions for which the particle 
positions are arranged in a unsmoothed HCP lattice are denoted with the 
suffix SPH, whereas all of the others adopt the smoothed density profile 
constructed according to the procedure described here.

Another issue which is found to have a significant impact on the ability of 
standard SPH to properly capture KH instabilities is the choice of the kernel
function. According to \cite{read10}, the accuracy of the momentum equation 
for particle $i$ is governed by the leading error 

 \be
    \vec {E_i}^{0}=\sum_j \frac{m_j}{\rho_j} \left[
  \frac{\rho_i}{\rho_j} +\frac{\rho_j}{\rho_i}\right]
   h_i \vec \nabla_i \bar W_{ij}~,
  \label{en0.eq}
  \ee

which vanishes in the continuum limit. However, this  does not hold
for a finite number of irregularly distributed particles or, more specifically, 
in the proximity of a contact discontinuity where a density step is present.
A possible solution consists of increasing the number of neighbors so as to 
improve kernel sampling, although this approach presents some difficulties 
when the commonly employed $M_4$ or cubic spline (CS) SPH kernel is used.
This occurs because  for a large number of neighbors the CS kernel is subject 
to the so-called clumping instability, in which pairs of particles with 
interparticle distance $q=r/h<2/3$ remain close together because the CS kernel 
gradient 
tends to zero below this threshold distance. A stability analysis 
\citep{mr96,borv04,pr05a,read10} 
shows that for the CS kernel, a Cartesian lattice of particles is unstable for 
$\eta \simeq 1.5 $ or $N_{sph}\sim 28,110$ when $D=2$ and $D=3$, respectively.
The clumping degrades the spatial resolution because it reduces the effective 
number of neighbors with which integrals are sampled, thus still having large
$\vec E_0$ errors even when the resolution is increased. 
To overcome this problem one can modify the kernel shape in order to have a 
non-zero kernel derivative at the origin. However, for a fixed number of 
neighbors, kernels of this kind have the drawback of giving a less 
accurate density estimate than that given by the CS kernel since near the 
origin the kernels have a steeper profile.

As an  alternative to the CS kernel, we consider here the linear quartic (LIQ)
 kernel, introduced by \cite{vrd10}, which is a quartic polynomial for 
$q_s\leq q\leq \zeta=2$ and is linear for $0\leq q <q_s$. 
The choice of the parameter 
$q_s$ determines the quality of density estimates. From a set of 2D Sod 
shock tube runs, \cite{vrd10} recommend $q_s=0.6$, which is the value adopted here.
For the functional form and normalization of the kernel see \cite{vrd10}.

Another kernel which has been introduced for the purpose of avoiding particle
clumping is the core triangle (CRT) kernel \citep{read10} which has constant 
first derivative for $0\leq q < \alpha$ and is similar to the CS kernel
for $\alpha\leq q \leq \zeta=2$.
This kernel is of the form

\be
w(q) = \frac{\sigma}{h^D} \left\{ \begin{array}{ll}
\left(-3\alpha+\frac{9}{4}\alpha^2\right)q +1+\frac{3}{2}\alpha^2-\frac{3}{2}\alpha^3 & 0 \le q < \alpha; \\
1-\frac{3}{2}q^2 +\frac{3}{4}q^3 & \alpha \le q < 1; \\
\frac{1}{4}\left(2-q\right)^3, & 1 \le q < 2; \\
0. & q \ge 2, 
\end{array} 
\right. 
\label{eq:ctspline}
\ee

where 
$1/\sigma=2\pi{\left(\frac{7}{20}+\frac{\alpha^4}{8}-
\frac{3}{20}\alpha^5 \right)},~4\pi\left(\frac{1}{4}+\frac{\alpha^5}{20}-
\frac{\alpha^6}{16} \right)$ for $D=2$ and $D=3$ , respectively. 
The value of $\alpha$ is fixed by the condition  of continuity for the 
second derivative, giving $\alpha=2/3$.
For the grid of initial conditions previously described, the sample of KH 
simulations is then constructed by performing  SPH runs with the same 
initial conditions  but using different kernels. We consider the CS kernel, 
together with the LIQ and CRT kernels. For all of these runs the number 
of neighbors is $N_{sph}=32 ~(\eta\simeq1.5)$. 

However, another solution for reducing sampling errors consists of 
still keeping a $B-$spline kernel function  but increasing its order 
\cite[sect.5.4]{pr12}. 
This approach presents the advantage of retaining the bell shape of the kernel, 
a feature which provides good density estimates \citep{fq96}. After the 
$M_4$ (CS) kernel, at the next order is the $M_5$ or quartic kernel

\be
w(q) = \frac{\sigma}{h^D} \left\{ \begin{array}{ll}
\left(\frac52 -q\right)^4 - 5\left(\frac32 -q\right)^4 + 10\left(\frac12-q\right)^4, & 0 \le q < \frac12; \\
\left(\frac52 -q\right)^4 - 5\left(\frac32 -q\right)^4, & \frac12 \le q < \frac32; \\
\left(\frac52 -q\right)^4, & \frac32 \le q < \frac52; \\
0. & q \ge \frac52, \end{array} \right. 
\label{eq:quarticspline}
\ee

which  is truncated to zero for $\zeta \geq 2.5$ and  has
 $\sigma={96}/{1199\pi} ,{1}/{20\pi}$ for $D=2$ and $D=3$,  respectively.

An additional set of SPH simulations was then performed using 
the $M_5$ spline as the kernel. For consistency with the other runs, we kept 
the same ratio of 
smoothing length to particle spacing ($\eta \sim 1.5$) so that for these runs 
the chosen number of neighbors is $N_{sph}=50$. A non trivial issue concerns the
 role of pairing instability for this class of kernels. Because the gradient 
of the $M_5$ kernel still goes to zero as $q\rightarrow 0$, one would expect 
the instability 
still to occur for $\eta \sim 1.5$. Nonetheless, it will be seen that KH 
simulations with the $M_5$ kernel do not exhibit particle clumping, in 
contrast with corresponding simulations performed with the CS kernel. 
This suggests that the stability properties of the $M_5$ kernel are better than 
those of the CS one; this topic will be discussed in a subsequent
dedicated section (\ref{khstab.sec}).

Table \ref{tab:cases} summarizes the main simulation parameters used in the 
KH tests.
For comparison purposes, a set of mirror runs was carried out  for the LIQ 
simulations in which Eq. (\ref{vsgv.eq}), for the signal velocity,
 was replaced in Eq. (\ref{duc.eq}) by Eq. (\ref{vspr.eq}) which is based on 
pressure discontinuities.

\subsubsection{Results}

Fig.~\ref{fig:vrdmap_1} shows the density maps at $t=\tau_{KH}$ for some of the
KH simulations listed in Table \ref{tab:cases}. The panels can be compared 
directly with the corresponding ones in Fig. 10 of \cite{vrd10}. A visual 
inspection reveals that the expected features of the KH runs are correctly 
reproduced and of a general consistency between the results produced by the two 
codes. In particular, the LP simulations which use the pressure-based AC signal 
velocities (\ref{vspr.eq}) appear to produce results almost identical to
the corresponding LIQ ones. Thus showing how, for the KH tests examined here,
the use of the two signal velocities can be considered equivalent, within the
spanned range of Mach numbers. 

Note, however, the absence of KH rolls  for the LP run with $M=0.4$, whereas 
for the same simulation in \cite{vrd10} the rolls have already been developed.
Given the general agreement between the two codes it is hard to ascertain the
origin of the discrepancy for this specific run.
However, the panels of Fig.~\ref{fig:vrdmap_1} show that the AC-SPH formulation,
and this point will be discussed in detail later in sect. \ref{RT.sec}, 
 is still unable to properly capture the development of KH instabilities for very subsonic shear flows.
This suggests how small differences in the time integration procedure 
of the two codes can become manifest in the long-term evolution of cold flows.

In Fig.~\ref{fig:vrdmap_2} the results for the same set of KH 
simulations of Table~\ref{tab:mach} are shown, but with the use  of 
kernels LIQ, CRT and $M_5$.
An important feature is now the appearance for the CRT and $M_5$ runs 
of KH rolls in the $M=0.4$ case. This improvement in the description of
 KH instabilities suggests that integration errors, that are present with the 
LIQ kernel, are now absent or reduced in the CRT and $M_5$ runs.
However, in the $M=0.2$ case, the kernels are still unable to capture the 
development of the KH instability.  For this test case, a high-resolution run 
($N=1024^2$) using the $M_5$ kernel (not shown here) still reveals the absence 
of any KH roll at $t=\tau_{KH}$, thus showing  that in SPH the problem of an 
accurate description of KH instabilities in the very subsonic regimes  is not
a resolution issue or due to the AC implementation. See also \cite{na11} for a 
recent discussion on this topic.

To test the effectiveness of the switch (\ref{salfa.eq}) in ensuring a fast
response of the $\alpha_i^C$ parameters to the presence of thermal energy
discontinuities, Fig.~\ref{fig:acmap}  renders the plots of the  AC 
parameters that are shown for the test runs of Fig.~\ref{fig:vrdmap_2}. 
The maps are color coded according to the range of values of $\alpha_i^C$.
The highest values lie in the range $\sim 0.7-0.8$ and are confined in a
narrow strip around the interface layers, with the floor value $\sim 0$ 
as the background value away from the discontinuities. 
Note that in general, and in particular for the $M=1$ test case, the maximum
 values of the $\alpha^C$ for the $M_5$ runs are below those of the other runs.
A result due to a faster growth rate of the instability, ensured by an
improved kernel accuracy in interpolating the variables.

\begin{figure*}
%\vspace{-3cm}
%\hspace{-0.5cm}
\centerline{
\includegraphics[width=15.2cm,height=7cm]{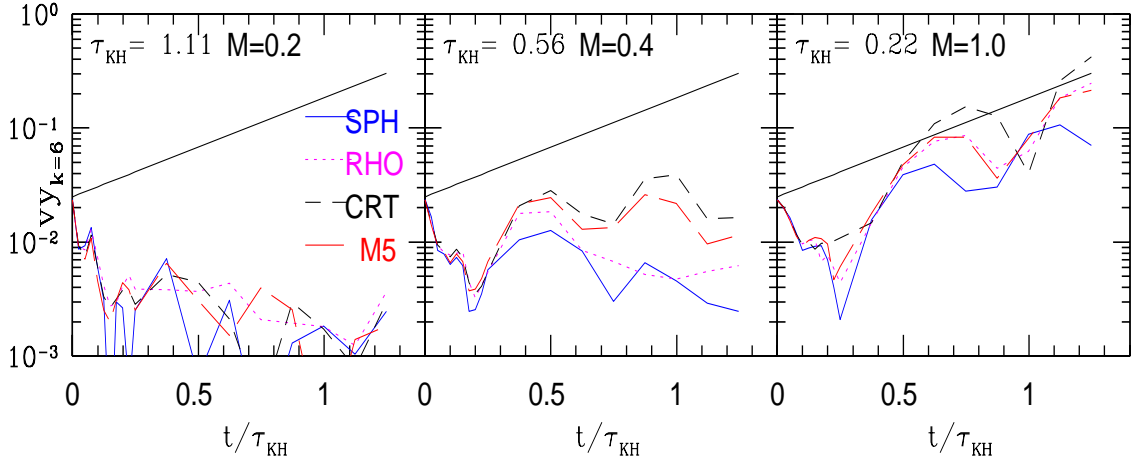}
}
\caption{Time evolution of the velocity field amplitude in the $y$ direction, 
as measured by the $k \lambda=12\pi$ mode of the Fourier transform of  $v_y$,
for some of the KH instability tests described in Sect. \ref {kh2d.sec}.
Each panel refers to KH simulations performed with the same Mach number, the
initial conditions set-up being given in Table \ref{tab:mach}. Within each 
panel, the  different curves are for AC-SPH runs with different simulation
parameters, as specified in Table \ref{tab:cases}. The black solid line 
indicates the expected linear theory growth rate.}
\label{fig:vymode}
\end{figure*}
The long term evolution of the KH tests is shown at $t=2$ 
in Fig.~\ref{fig:vrdmap_3}. The overall features of the density plots for 
different runs are similar to their counterparts displayed in Fig. 11 of
\cite{vrd10}. Note the tendency for the $M_5$ runs to the appearance of 
small scale features at the layer contacts.

The aim of this section is to test the consistency of the present AC implementation
by comparing results, extracted from a suite of AC-SPH simulations of the 2D KH
instability problem, with those of similar runs \citep{vrd10}.
The results of the KH tests indicate a code behavior which is in accord with what 
expected and with the simulations of \cite{vrd10}.

However, \cite{vrd10} argue that it is not the absence of the AC term the 
main reason of the SPH failure to develop KH instabilities, although the presence
of AC is necessary for the long-term evolution of the instabilities, but this
difficulty of SPH is rather due to two distinct reasons. 
The first issue is a general 
problem of consistency of the initial condition set-up, as SPH by definition can 
only deal with smoothed quantities and therefore its application to problems 
where sharp discontinuities are present leads to inconsistencies. This is part
of the more general problem \citep{rb10} of properly smoothing in numerical
simulations of hydrodynamic test problems the discontinuities initially present 
at the interfaces, so to achieve convergence in the solution. This aspect of the KH
test problem can be cured by properly stretching the initial particle lattice 
so as to introduce a smooth density transition at the interfaces. The results 
indicate a significant improvement in the capability of SPH to properly capture
the correct growth rate of the KH instability.  

The other issue which causes a poor performance  of SPH when handling KH 
instabilities is the leading error in the momentum equation due to incomplete 
kernel sampling. This error can be reduced by removing 
particle clumping, which depends on the kernel stability properties.
In fact, the kernels LIQ and CRT have been introduced \citep{read10,vrd10}  
with the aim of removing the clumping instability. Since the results of the 
simulations suggest performances  for the $M_5$ kernel that are quite similar 
to those
achieved by these kernels, it is therefore interesting to better quantify the 
relative performances of these kernels.

Following \cite{ju10}, we then measure the growth rate of 
the KH instability for some of the runs and compare it with the linear 
theory growth rate expectation
$\propto e^{t/\tau_{KH}}$. This is achieved, using Fourier 
transforms, by measuring the time-evolution of the $\lambda=1/6$ growing mode of the $v_y$ 
velocity perturbation component. The details of the procedure are given in 
\cite{ju10} and will not be reported here. For the sake of clarity,
 the results of 
the LIQ runs are not displayed since they are quite similar to those of the CRT 
simulations. Moreover, we only display the growth rates for three different KH
test cases, those with Mach number $M=0.2,0.4$ and $M=1$, the results of the 
others being intermediate between these.

There are a number of distinct features which are apparent from the panels of 
Fig. \ref{fig:vymode}. The first is that simulations with smoothed IC ( RHO) 
perform systematically better than the unsmoothed ones (SPH). The second is that
only for $M=1$ the simulations with kernels CRT and $M_5$ are able to correctly
recover the expected growth rate. Finally, this capability degrades progressively 
as one considers lower Mach numbers.
While this behavior is in accordance with the visual impression of the maps previously 
displayed, and confirms that the present SPH implementation still has
problems in the description of instabilities in subsonic flows, it is 
interesting to note that the performances of the $M_5$ runs are similar to those
 obtained using the CRT kernel.
This behavior is particularly interesting, since the simulations have been 
performed  setting the ratio of the smoothing lengths to particle spacing 
to $\eta\simeq1.5$ so that the pairing instability, that is present in the runs
which use the CS kernel, should be present in the $M_5$ simulations 
as well.

\begin{figure*}
%\vspace{-3cm}
\centerline{
\includegraphics[width=17.2cm,height=10.2cm]{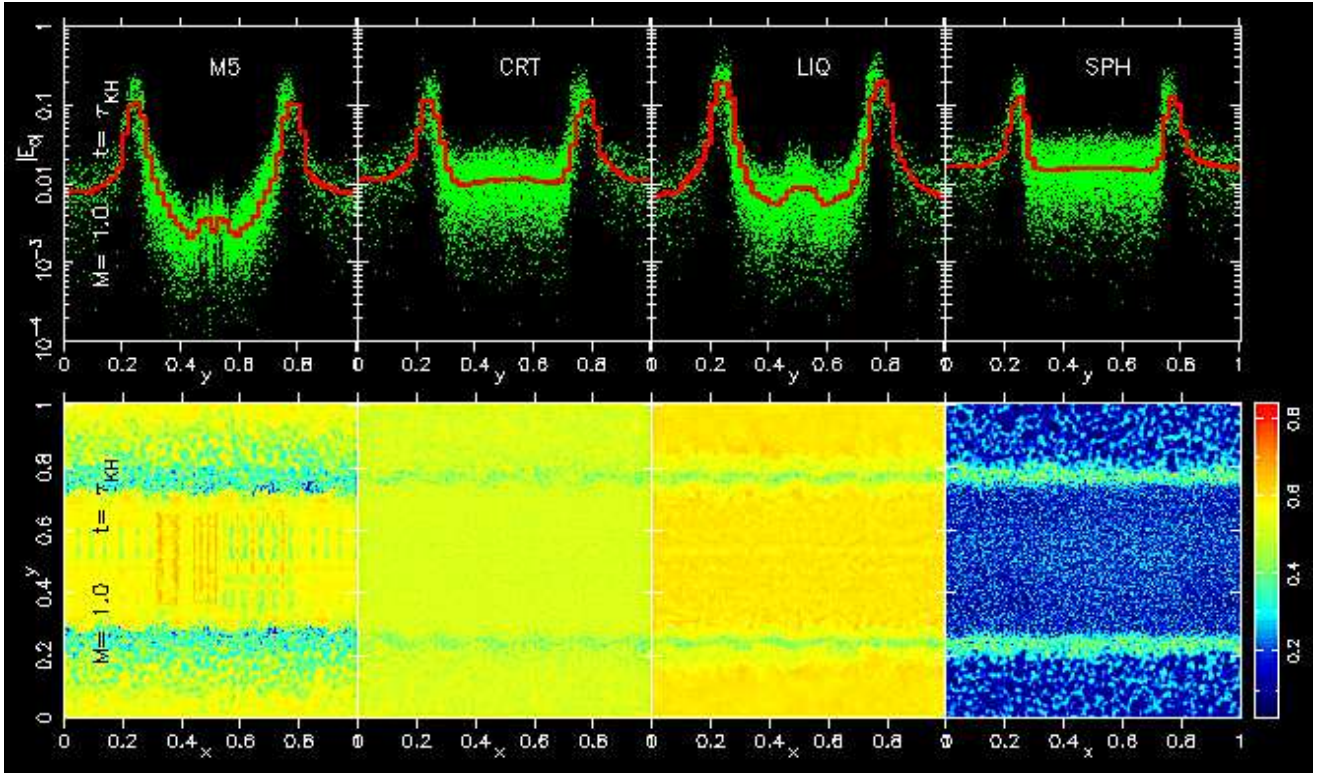}
}
\caption{ Top panels: distribution at time $t=\tau_{KH}$
of the errors $|\vec {E_i}^{0}|$ plotted versus  $y$, as defined 
by Eq. (\ref{en0.eq}), 
for the  KH runs with Mach number $M=1$ and different simulation
parameters (See Table \ref{tab:cases}). The red histograms show the mean values of
the binned distributions.
Bottom panels: each panel shows the nearest neighbor map of the 
 run in the corresponding   top panel. This is defined as the distribution, 
interpolated at the map grid points of the  normalized 
distances $q_i^2=\delta_i^2/h_i$, where $\delta_i^k$ is the distance 
$|\vec x_i-\vec x_k|$
of the $k-$th neighbor of the particle $i$ and the neighbors are sorted 
so that $\delta_i^k<\delta_i^{k+1}$ .
}
\label{fig:errmap}
\end{figure*}

How the  clumping  instability affects sampling errors can be assessed by 
computing the particle errors $\vec E_i^0$, according to Eq.~(\ref{en0.eq}). 
The distribution, throughout the simulation domain of the $\vec E_i^0$ errors 
versus $y$, is shown in the top panels of Fig.~\ref{fig:errmap} 
for those $M=1$ simulations 
performed using different kernels. The solid lines represent the mean of the 
binned distributions. Similar plots have been produced by \cite{vrd10} and
their Fig.~3 can be used for comparative purposes. 

As expected, the largest  $\vec E^0$  errors are present in the SPH simulation, 
whereas better results are obtained by using the LIQ and CRT kernels, 
as indicated by the error distribution in their respective panels. This is a result
 of the absence of particle clumping for these simulations, owing to the 
specific stability properties of these kernels \citep{read10,vrd10}.
 
A striking feature is given vice versa by the $\vec E^0$  error distribution 
of the simulation performed using the $M_5$ kernel, which, in fact,  shows how 
 the magnitude of the $\vec E^0$  errors are even smaller than 
those of the two runs CRT and LIQ for this kernel. 
A result which is in accordance with what has been found 
previously by analyzing the growth rates, suggesting that, since all of the 
simulations were performed using the same $\eta$, the stability properties 
of the $M_5$ kernel are better than those of the CS one.

To further investigate this point, the bottom panels of Fig.~\ref{fig:errmap} 
show the `nearest neighbor' map of the corresponding top panels. This is 
defined by interpolating the quantity $q_i^2$ at  the grid points, 
according to the SPH prescription, where for the $k-th$ neighbor 
$q_i^k=|\vec x_i-\vec x_k|/h_i$ and the neighbors have been sorted according to
their distance from the particle $i$ itself.

The map of the  $M_5$ kernel indicates a distribution of the second nearest
neighbor distribution $q_i^2$ that is quite similar to that displayed by 
the CRT and LIQ kernels. 
The only difference being at the layer interfaces where 
  the distribution of quantities $q_i^2$ for 
the $M_5$ kernel is slightly 
shifted towards smaller values than for the other kernels.
Note, vice versa, how for the CS kernel because of the pairing instability, the
distribution of the neighboring distances is flipped with respect that 
of the other kernels.

The results of Fig.~\ref{fig:errmap}, however, clearly show the absence of 
clumping instability for the $M_5$ kernel. In the next section, we  
investigate, in more detail, the stability properties of this kernel.

\begin{figure*}
\centerline{
\includegraphics[width=15.2cm,height=8cm]{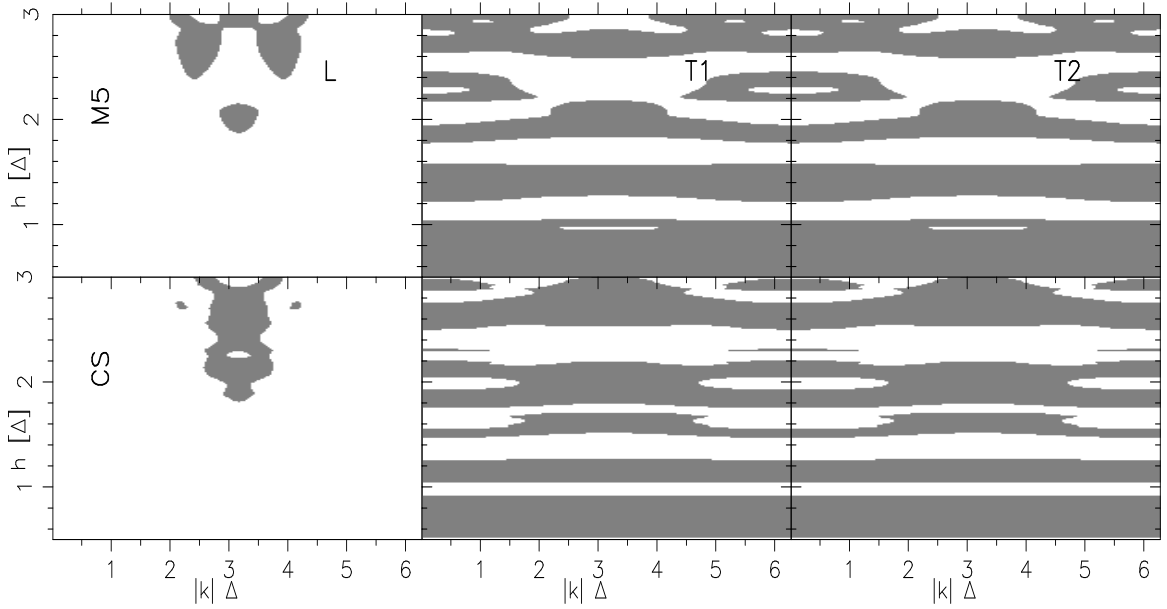}
}
\caption{Contour plots of the dispersion relation (\ref{omega.eq}) are
shown as a function of wavenumber $k$ and smoothing length $h$ for a
regular lattice of particles with spacing $\Delta$, we consider a  wavevector 
with orientation $\vec k=k(1,0,0)$. 
Gray areas indicate the 
instability regions for which $\omega^2<0$. From the left to right, the top 
panels show in the case of the $M_5$ kernel the instability regions of the 
longitudinal and the two transverse waves. 
The bottom panels refer to the CS ($M_4$) kernel.}
\label{fig:stabw}
\end{figure*}

\subsubsection{Stability issues}
\label{khstab.sec}
The stability properties of SPH have been investigated by a number of authors
 \citep{sw95,mr96,mo00,borv04,read10,de12}. The instabilities are studied analytically
 by analyzing the dispersion relation for sound waves of
small amplitude, propagating in a uniform medium. We now derive the dispersion 
relation for the SPH equation of motion in a manner similar to the analysis of 
 previous authors.
 A uniform lattice of equal mass particles of mass 
$m$, density $\rho_0$, pressure $P_0$ and sound speed $c_s^2=\gamma P_0/\rho_0$ 
is perturbed by a small wave 

\be
\vec {x_i} = \vec {x_i}^0+\vec a \exp[\vec k \cdot\vec {x_i}^0 -\omega t]~,
\label{xdis.eq}
\ee
where $\vec a$ is the perturbation, $\vec k$ the wavevector and 
$\vec {x_i}^0$ are the unperturbed particle positions.
By linearizing the equation of motion for the perturbation one has the 
dispersion relation

 \begin{eqnarray*}
\lefteqn{
  \omega^{2} a_{\mu}=\left [ \frac{2 mP_0}{\rho^2_0}
\sum_j H_{\mu \nu} (1-cos(\vec k \cdot\vec {x_{ij}}^0) a_{\nu}+
\right.}  \\
   & & \mbox{}+\left.(\gamma-2)\frac{m^2P_0}{\rho_0^3} (\vec a \cdot \vec q^i ) 
 q^i_{\mu}\right]~,
  \label{omega.eq}
 \end{eqnarray*}

where  the summations are over the neighbors $j$ of  
particle $i$, the vector $\vec q_i$ is defined as 

\be
\vec {q_i} = \sum_j \sin \vec k \cdot\vec {x_{ij}}^0
   \vec \nabla_i W_{ij}~,
\label{qdis.eq}
\ee

and ${\bf H}(W)$ is the Hessian of the kernel
\be
H_{\mu \nu}=\frac{\partial^2 W(r)}{\partial x_{\mu} \partial x_{\nu}}.
\label{hess.eq}
\ee

For a given smoothing length $h$ and wavevector $\vec k$, the stable
solutions  of Eq.~(\ref{omega.eq}) are defined by the condition 
$\omega^2\geq 0$ . Solutions for which 
 $\omega^2< 0$  represent exponentially growing or decaying perturbations.
Moreover, it is useful to define a numerical sound speed $C^2_{num}=\omega^2/k^2$ 
and a scaled numerical sound  speed $c^2_{num}=C^2_{num}/c^2_s$. Clearly, the 
condition $c^2_{num}=1$ should be satisfied to correctly model the 
sound speed.

We now examine the stability properties of the CS and $M_5$ kernels. 
For simplicity, we consider a plane wave 
propagating along the $x-$axis, $\vec k=k(1,0,0)$, and assume $\gamma=5/3$.
The bottom panels of Fig~\ref{fig:stabw} show, for the longitudinal and the two 
transverse waves of the perturbation, the instability regions of the 
CS kernel, these are denoted by the gray areas and represent the solutions 
to Eq.~(\ref{omega.eq}) , in the domain $(h,k)$, for which $\omega^2< 0$. 
Similarly, the top panels are for the $M_5$ kernel.

The longitudinal wave perturbation is responsible for the clumping instability, 
whereas the traverse waves give rise to the so-called banding instability 
\citep{read10}. Unlike the clumping instability, the latter is relatively 
unimportant in causing sampling errors \citep{read10} and will not be further 
considered here. A comparison of the two stability plots for the longitudinal
wave solution clearly shows that the stability properties of the $M_5$ kernel
are much better than those of the CS one. 

This is part of a more general 
result which was already recognized by \citep{mr96}: the stability properties of
SPH are improved, as higher order spline kernels are used in place of the CS kernel.
This occurs, basically, because the higher the order of the spline, 
the better it
approximates a Gaussian kernel. 
%This is characterized by a Fourier transform 
%which, for a given $h$, falls off with $k$ faster than any  kernel with
%compact support.
In Eq.~(\ref{omega.eq}) one can see that the numerical sound speed 
 $c^2_{num}$  depends on the first and second derivative of kernel $W$.
Ideally, one should have $c^2_{num}\simeq 1$ to accurately describe the sound 
wave propagation and this condition is fulfilled as smoother kernels 
are used, so as to keep the numerical dependency of $c^2_{num}$ as weak 
as possible. 

Note however that here, differently from the stability properties of 
the CRT kernel, the clumping instability is not entirely suppressed but 
rather it is 
present whenever $\eta \simgt 2.5$.

Finally, it must be stressed that the paring instability that occurs in the 
hydrodynamic tests described here, is unlikely to have a significant impact 
on many astrophysical problems of interest, where very cold flows are absent.
\cite{ra99} estimate for instance, from SPH simulations of a stationary fluid,
that lattice effects become important when velocity dispersions are
below $\sim3\% -4\%$ the sound speed.

 The results of this section and of the previous one, therefore, suggest that 
in order to avoid the pairing instability,  the $M_5$ kernel can be considered 
a viable alternative to the use of the otherwise steeper CRT and LIQ kernels,
provided that the parameter $\eta$ is consistently rescaled.
 In the next section, we then discuss the related performances of these 
kernels in a test simulation.

\subsection{Sod's shock tube}
\label{sod.sec}
A classic test used to investigate the hydrodynamic capabilities of
SPH codes is the  Sod shock tube problem 
\citep{hk89,wa04,sp05,pr08,ta08,ro09}. This test consists in a fluid,
 initially at rest, in which a membrane located at $x=0$ separates the fluid on the right, 
with high density and pressure, from the fluid on the left, with lower 
density and pressure. The membrane is removed at $t=0$ and a shock wave develops
propagating toward the left, followed by a contact discontinuity and a 
rarefaction wave propagating to the right. 

A well known problem of standard SPH codes in reproducing the analytic solution of 
the shock tube problem is the presence of a pressure discontinuity that arises
across the propagating contact discontinuity.
Simulations incorporating an artificial thermal conductivity term in the SPH 
equations \citep{pr08,ro09,pr12} show shock profiles in which density 
and thermal energy are resolved across the discontinuity,
  hence giving a continuous pressure profile.
However, in these runs the AC formulation adopts the AC signal velocity 
(\ref{vspr.eq}) where jumps in thermal energy are smoothed in accordance 
with the presence of pressure discontinuities.
This is in contrast with the AC signal velocity (\ref{vsgv.eq}) employed here,
for which in absence of shear flows, contact discontinuities are unaffected
and therefore cannot be used to remove the blip seen at the contact discontinuity 
 in the pressure profile of the shock tube SPH runs.

However, the application of the AC-SPH code to this test 
is nonetheless of interest because it can still be used 
to validate  code performances.
In particular, we consider a 3D setup of the shock tube test and with these
initial conditions we construct a set of AC-SPH simulations performed with 
different kernels and neighbor numbers. Shock tube profiles extracted 
from these simulations are compared with the aim of assessing the goodness 
of different kernels in reproducing, for a given test problem,
 profiles of known analytic solutions.

\begin{figure*}
\centering
\includegraphics[width=15.2cm,height=12.2cm]{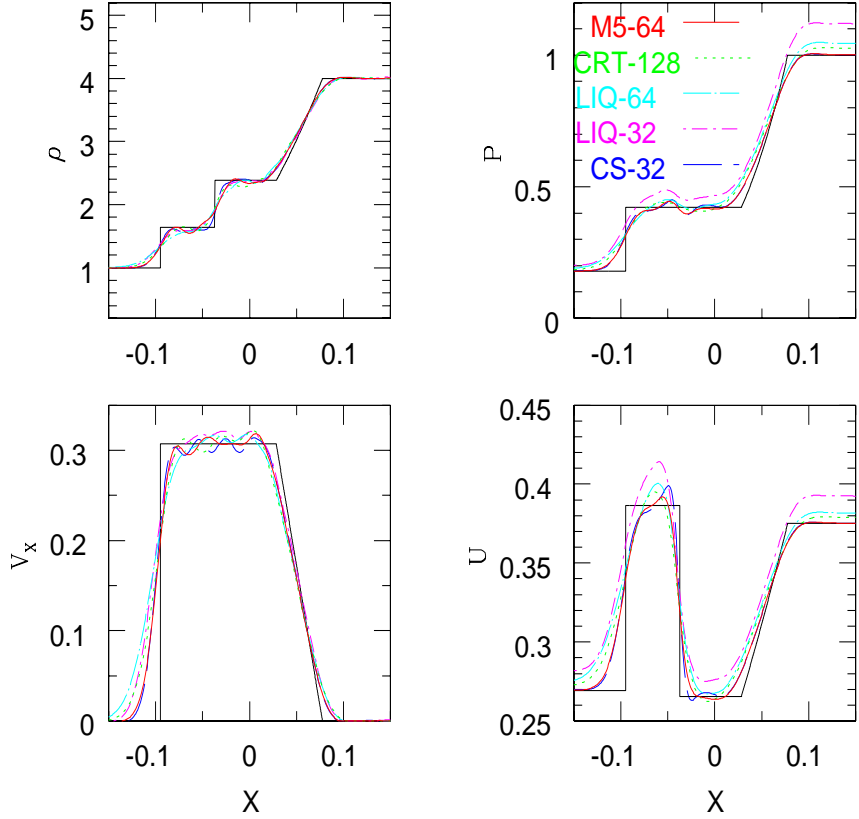}
\caption{Results at $t=0.12$ of the 3D shock tube test. 
The profiles of: density, pressure, thermal 
energy and velocity  are plotted clockwise from top left.
 The solid black line represents the analytical solution, 
while lines with different styles and colors are the profiles of the AC-SPH runs
with different kernels and neighbor numbers, as illustrated in the pressure
panel.}
\label{fig:tubec}
\end{figure*}

The initial condition setup consists of an ideal fluid with $\gamma=5/3$, 
 initially at rest at $t=0$. An interface set at the origin separates the  fluid
on the right with density and pressure $(\rho_1,P_1)= (4, 1)$ from the
fluid on the left with $(\rho_2,P_2)=(1, 0.1795)$. 
To construct these initial conditions, a cubic box 
of side-length unity was filled with  $10^6$ equal mass particles, so 
that $800,000$ were placed in the right-half
 of the cube and $200,000$ in the left-half. 
The particles were extracted from two independent
uniform glass-like distributions contained in a unit box consisting of
  $1.6 \cdot 10^6$ and $4\cdot 10^5$ particles, respectively.
This initial condition setup is the same previously implemented 
in Sect. 5.1 of V11, to test the time-dependent AV scheme 
described in Sect. \ref{avvis.sec}.

For the same initial setup, to investigate the performances of different 
kernels in reproducing the 
analytic profiles of the shock tube problem,  
we perform runs with different kernels and neighbor numbers. The kernels 
considered are : CS, LIQ, CRT and $M5$. The number of neighbors varies 
in power of two between $32$ and $128$.
 The SPH runs were realized by imposing periodic boundary conditions along 
the $y$ and $z$ axes and the results were examined at $t=0.12$.
% has the effect of removing the pressure blip
We show results obtained using the standard AV formulation, the results of
the other runs being unimportant from the viewpoint of kernel performances.
 
Some of the simulation profiles extracted 
from the 3D SPH runs of the shock tube test are shown in Fig.~\ref{fig:tubec}.
 A striking feature is the 
wide scatter between the pressure profiles of the runs performed using 
different kernels or neighbor numbers. The same behavior is present for the
thermal energy profiles, 
whilst very similar density profiles are exhibited by the same runs. 

There are several conclusions that can be drawn from the results of 
Fig.~\ref{fig:tubec}. The performances of the $M_5$-64 run are quite 
similar to those of CS-32, although for the former simulation a closer 
inspection reveals a slightly 
better treatment of the thermal energy spike at the contact 
discontinuity, the spike being due to the initial condition set-up. 

The worst results are obtained by the LIQ simulations when using  $N_{sph}=32$
or $N_{sph}=64$ neighbors. The profiles of the LIQ-128 run are not shown here
 to avoid overcrowding in the plots, and are quite similar to those of  
simulation CRT-128.
This clearly demonstrates the need for this class of kernels to use a large 
( say $\simgt 128$ ) number of neighbors, so as to compensate for the density 
underestimate due to the steeper profiles introduced to avoid the pairing 
instability.

To better quantify this density bias, Table ~\ref{tab:rhotest} reports, for 
different kernels and neighbor number, the mean SPH density estimated
from  a glass-like  configuration of $N=10^6$ particles, in a unit periodic box 
of total mass one.
The results illustrate how the density estimate of the $M_5$ kernel with 64
neighbors is comparable to the CS one obtained using 32 neighbors, the 
value of $\eta$ being the same ($\eta\simeq 1$). Vice versa, for the LIQ and
 CRT 
kernels, only when $N_{sph}=128$ does the mean density  approach the $M_5$-64 
estimate. 

The density values given in Table ~\ref{tab:rhotest} also help to explain
the rather poor performances of the LIQ kernel when using 32 neighbors.
From Fig.~\ref{fig:tubec} it can be seen that for the corresponding run
the relative  pressure error is $\varepsilon_P\simeq 10\%$ in the unperturbed
right zone of the cube.
This error is already present in the pressure profile at $t=0$ and it is
originated from the density error due to the adopted kernel and neighbor 
number, together with the use of an entropy-conserving code to perform
the simulations.
Initially, particle entropies are assigned according to the initial conditions 
so that, for particles satisfying $x_i>0$, 
$A_i\equiv A_1=P_1/\rho_1^{\gamma}\simeq0.1$.
During the integration particle pressures are calculated according to
$P_i=A_i \rho_i^{\gamma}$, and for $x>0$ a relative pressure error
 $\varepsilon_P\simeq \gamma \varepsilon_{\rho}\simeq 10\%$ 
is present when $\varepsilon_{\rho}\simeq 6\%$.

Taken at face value, the results of Table ~\ref{tab:rhotest} 
demonstrate that in 3D simulations a conservative lower limit for the kernels 
with a modified shape, should be to assume at least $N_{sph}\simgt 200$
neighbors. In fact, in a recent paper \cite{read12} presented a new 
formulation of SPH where they adopted, as reference, the so-called 
high-order core triangle (HOCT, \cite{pr05a}). The profile of this
 kernel is a generalization of that of the CRT kernel and the authors
assume $N_{sph}=442$ as the reference value for their scheme.

However, the results of the previous sections suggest that the stability 
properties of the $M_5$ kernel can be profitably used, with an appropriate 
choice of the $\eta$ parameter, to avoid the pairing instability when 
dealing with tests of hydrodynamic instabilities in which cold flows are 
present.
 
Finally, the density estimates of Table ~\ref{tab:rhotest} suggest that 
great care should be used when deciding on the goodness of a particular kernel 
on the basis of its relative performances in terms of density estimates.
The results of the  3D Sod shock tube clearly indicate how there could be
a large difference between the simulation and the expected solution profile 
of some hydrodynamic variable, such as pressure or thermal energy, 
while having, at the same time, a much smaller  difference in the corresponding 
density profile.

\subsection{Sedov blast wave}
\label{sedov.sec}

The Sedov blast wave test is used to validate, in three dimensions, the code 
capability in the strong shock regime. The test consists of a certain amount 
of energy $E$ which is injected at $t=0$ into a very small volume of an 
ambient medium of uniform density $\rho$. The spherically symmetric shock
propagates outward from the initial volume and at  the time $t$ the 
shock front is located  at the radius \citep{sd59}

\be
R(t)\simeq \beta (E t^2 /\rho) ^{1/5}~,
\label{rsedov.eq}
\ee
where $\beta \sim 1.15$ for $\gamma=5/3$.

Previous investigations \citep{rp07,me10} showed that,
owing to the large discontinuities initially present in the thermal energy, 
incorporating an artificial conduction term in the energy equation greatly 
improves the description of the shock front in the simulations.
Without the presence of this term, the initially large discontinuity in
the thermal energy will soon give rise to a disordered particle distribution 
thus degrading the shock profile \citep{rp07}.

The initial setting of the test is realized as follows. A HCP lattice of 
$N=2\cdot 64^3$ 
equal mass particles is arranged in a cubic box of side length unity. The particle
masses are chosen so as to give $\rho=1$ and periodic boundary conditions are
imposed. The nearest particle to the position $\{0.5,0.5,0.5\}$ is chosen as 
the center particle. 
In order to consistently represent  a point-like explosion with the given 
numerical resolution, the particles $j$ comprised within the
kernel radius $\zeta h_i$ of the center particle $i$  are given an initial thermal
 energy such that the total injected energy is $E=1$. This blast wave energy 
is distributed among the neighboring particles not uniformly but with a weight 
proportional to $W_{ij}$.
\begin{table}
\caption{ Average densities and sample standard deviations estimated from the 
SPH densities of a configuration of one million particles. These are 
arranged in a 
glass-like configuration inside a cubic periodic box of side length unity 
and total mass one. The SPH particle densities have been calculated for a 
variety of kernels and neighbor numbers $N_{sph}$ (see text).}
\def\arraystretch{1.1}
\begin{tabular}{l|ccc}
\hline
\hline
&\multicolumn{3}{c}{$ N_{sph}$}  
\\ 
   \textsf{Kernel}  &   $32$   &   $64$   &
 $128$    \\
\hline
 \textsf{CS}  & 1.005 $\pm$ 0.035   & 1.002 $\pm$ 0.019    & 
1.0018 $\pm$ 0.012   \\
\textsf{LIQ}  & 1.06 $\pm$ 0.028   &  1.024 $\pm$ 0.016    & 
1.01 $\pm$ 0.011   \\
\textsf{CRT}  & 1.079 $\pm$ 0.039   &  1.035 $\pm$ 0.020    & 
1.016 $\pm$ 0.012   \\
\textsf{M5}  &  1.022 $\pm$ 0.052  &  1.003 $\pm$ 0.024    & 
1.0008 $\pm$ 0.015   \\
\hline
\end{tabular}
\label{tab:rhotest}
\end{table}

Using this initial condition set-up, we ran three test simulations, which differ 
in the choice of the adopted kernel and neighbor number.
For the CS kernel, we use $N_{sph}=64$  neighbors. We also 
run two other test cases, now using the CRT and $M_5$ kernel and $N_{sph}=128$ 
 neighbors. All of the simulations were performed using the implemented AC
scheme and the time-dependent  AV formulation with 
parameters given by $\{ \alpha_{min}, \alpha_{max}, l_d\}=\{0.1,1.5,0.2\}$ .

\begin{figure}
\hspace{-0.7cm}
\includegraphics[width=10.2cm]{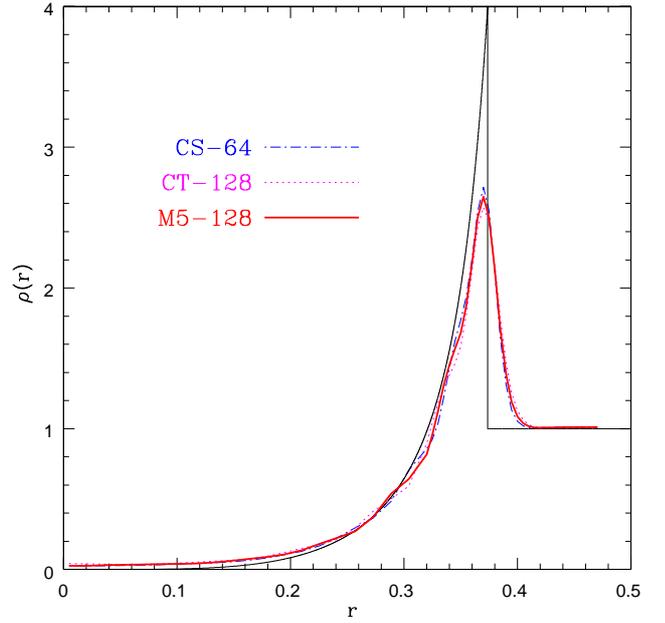}
\caption{Radial density profiles of the 3D Sedov blast wave test at $t=0.06$.
The solid black line indicates the analytic solution, while the simulation
profiles are obtained by averaging for each radial bin SPH densities 
calculated from the particle distributions over a set of grid points 
located at the surface of a spherical shell and uniformly spaced in 
angular coordinates.
All of the runs were performed using $N=524,288$ particles.}
\label{fig:sedov}
\end{figure}

Fig.~\ref{fig:sedov} shows the radial density profiles at $t=0.06$ for the 
different test runs. The solid black line represents the expected analytic solution,
with the shock front being located at $r\simeq 0.37$.
Radial simulation profiles were obtained by averaging, for each radial bin 
$r_k$, SPH densities calculated from the particle distributions over a set 
of $(\theta,\phi)=(20,20)$ grid points uniformly spaced in angular coordinates:
these are located at the surface of a sphere with radius $r_k$. 
The radial spacing is not 
uniform but is chosen so as to guarantee an accurate sampling of density in
the proximity of the shock front, with about $\sim 40$ radial bins between 
$r\sim0.34$ and $r\sim0.42$. 

All of the simulations are in fair agreement with the analytical solution and 
the differences between the simulation profiles are negligible.
At the shock front, the profiles exhibit a density jump of about  $\sim 2$, whereas 
the analytical solution gives a compression factor of $\gamma+1/\gamma-1=4$.
These results are in accordance with previous findings \citep{rp07,sp10b,hs10} and
indicate that for 3D SPH simulations of the  Sedov-Taylor point explosion problem
the simulation profiles converge to the analytical solution as the
resolution is increased, with approximately $\sim 345^3$ particles \citep{rp07} 
being required to fully resolve the shock front. 

Fig.~\ref{fig:sedov} can also be used to verify the behavior of the individual 
time-step algorithm. For problems involving very strong shocks, as 
demonstrated by \cite{sm09}, individual particle time-steps
 must be properly restricted so to avoid that in the proximity of the shock front
they do not satisfy the local Courant condition, thus leading to inaccuracies 
in the integration. In fact, SPH simulations using individual time-steps, but 
without an appropriate limiter, will fail to predict the expected solution
profile (see Fig. 3 of \cite{sm09}). 
The profiles of Fig.~\ref{fig:sedov} demonstrate the correctness of the algorithm 
used to update the time-steps, although different from that devised by \cite{sm09}.

The present scheme adopts individual particle time-steps $\Delta t_i$  that
 can vary in power-of-two subdivisions of the largest allowed 
time-step $\Delta t_0 \geq \Delta t_i$ \citep{hk89}.
At each step, particles whose time bin is synchronized with the 
current time are defined as `active' and their hydrodynamic quantities, as well 
as their smoothing lengths and densities are consistently updated.
However, non-active particles $j$, that are neighbors of  an active 
particle $i$ are defined as `low-order active' particles and their 
hydrodynamic variables, as well as their time-step constraints but not their 
forces, are recalculated.
This criterion is applied regardless of the local shock conditions and
no particular conditions are imposed on $\Delta t_j$ which depend on
$\Delta t_i$.

\subsection{The blob test}
\label{hotb.sec}

The blob test is another hydrodynamic test where results of standard SPH
differ significantly from those produced by grid based simulations 
\citep{ag07,read10,ch10,hs10,mu11}. The test consists of a gas cloud of radius 
$R_{cl}$ placed in an external medium ten times hotter and less dense than the 
cloud, so as to satisfy the pressure equilibrium. A large enough wind velocity 
$v_{W}$ is given to the hot low-density medium so that a strong shock 
wave strikes the cloud. The interaction of the cloud with the supersonic 
medium produces a number of effects that are of interest in an astrophysical 
context, such as gas stripping and fragmentation.

Initially, the blob is perturbed by the development of Richtmyer-Meshkov 
and Rayleigh-Taylor instabilities \citep{ag07}. After wards, large-scale 
($\sim R_{cl}$) KH instabilities are created at the cloud surface because 
of the shear flows due to the supersonic wind. This non-linear phase 
is supposed to develop over a time-scale 
$\tau_{cr}\sim 2 R_{cl}\sqrt{\chi}/v_W$ \citep{ag07}, where $\chi$ is the density 
contrast, after which cloud disruption will take place.

In order to investigate the capability of the AC-SPH code to properly follow the
hydrodynamic of the blob test, we compare results extracted from a set of SPH
simulations realized with the same initial conditions but with different 
numerical parameters. The numerical setup of the test is the same as in 
\cite{read10}, to which we refer for more details. A spherical cloud of
radius $R_{cl}=197kpc$ is placed in a periodic, rectangular box  of size
$\{L_x,L_y,L_z\}=\{2,2,8\}Mpc$. The cloud has density 
$\rho_{cl}=4.74 \times 10^{-33} gr cm^{-3} = \chi \rho_{ext}$ and temperature 
$T_{cl}=10^6 K=T_{ext}/\chi$. The ambient medium has density and temperature 
so that $\chi=10$. The cloud is initially located at $\{1,1,1\}Mpc$ and 
the ambient medium is given a wind velocity $v_W=1,000 km s^{-1}$, so that
 for an adiabatic index $\gamma=5/3$ its Mach number is $M=2.7$. 
An HCP lattice of equal mass particle is 
constructed in order to satisfy the above density requirements so as to use for this
version of the test, a total number of $N\sim1.1 \times 10^6$ particles, as
in \cite{hs10}.
Finally, a velocity perturbation is imposed at the cloud 
surface in order
to trigger the development of an instability; amplitude  and modes are given 
in appendix B of \cite{read10}. 

We compare results from SPH simulations 
where three different spline kernels have been used : CS, CRT and $M_5$.
We use $128$ neighbors for the simulations with the CS and CRT kernels and
$220$ for the run with the $M_5$ kernel, so that the ratio $\eta$ is the 
same for all the runs ($\eta \sim 1.5$). For the CS kernel, we run a 
simulation where the AC term (\ref{duc.eq}) is absent in the equation of 
thermal energy evolution (CS-128 NOAC), and three other simulations
 (CS-128, CRT-128, M5-220) in which the AC term 
(\ref{duc.eq}) is incorporated in the energy evolution equation.
In the following, the simulation AV parameters are set as follows: 
$\{ \alpha_{min}, \alpha_{max}, l_d\}=\{0.1,1.5,0.2\}$ 

\begin{figure*}
\centerline{
\includegraphics[width=14cm,height=11.3cm]{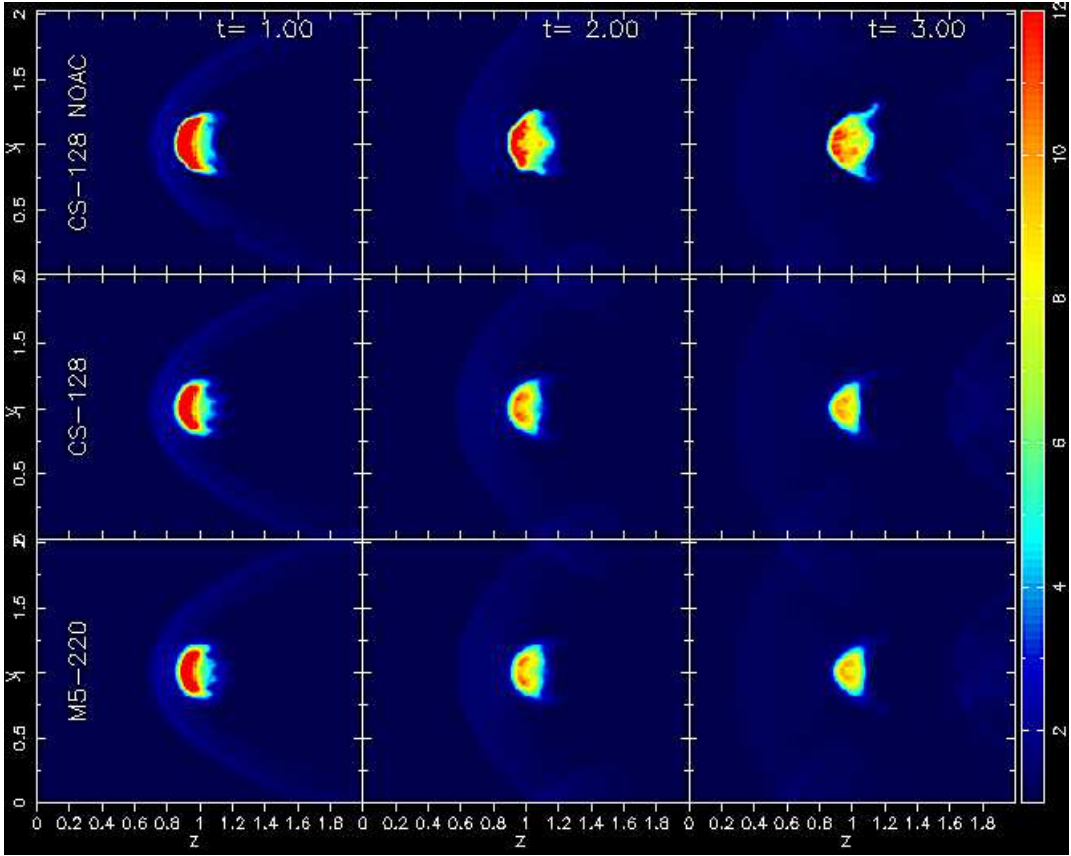}
}
\caption{  Density maps of the blob test in the central plane $x=L_x/2$ at
$t=1,2,3$ for SPH runs with (CS,~M5) and without (CS NOAC) the AC term. Time 
is in units of $\tau_{KH} \sim 1.6 \tau_{cr}$ and the number accompanying 
the kernel label indicates the number of neighbors of the run.
 Axis units are in $Mpc$.}
\label{fig:blobmap}
\end{figure*}

For several runs Fig.~\ref{fig:blobmap} shows the gas density maps at three
different times: $t=1,2,3$; the time being in units of 
$\tau_{KH} \sim 1.6 \tau_{cr}$ \citep{ag07}.
The maps have been evaluated according to the SPH prescription on a 2D
grid of $(200x800)$ points in the central $yz$ plane located at $x=L_x/2$.
The top panels of Fig.~\ref{fig:blobmap} are for the standard NOAC run 
CS-128. It can be seen that in this case, unlike what was found in 
mesh-based simulations \citep{ag07}, the cloud survives the impact of 
the supersonic wind striking its surface and there is no disruption. 
This is due to the absence of fluid instabilities developing 
at the cloud surface.  If they were present  they would in turn lead to the 
stripping of material and to the cloud break-up.
This code behavior is similar to what was found in previous findings 
\citep{ag07,read10,hs10}.

On the contrary, incorporating the AC diffusion term in the SPH equations 
leads to a significant improvement in the code capability to 
properly model the cloud evolution. This can be seen from the middle and bottom 
row panels of Fig.~\ref{fig:blobmap}, in which the density maps 
are displayed  for the CS-128 and M5-220 runs. The simulation performed using the 
CRT kernel is not shown since its performances are quite similar to those 
obtained using the CS kernel.

This behavior is consistent with the expectation that introducing AC removes 
the  numerical effects which in SPH prevent the treatment of
  contact discontinuities when large density jumps are present,
 and thus the inconsistencies that suppress the growth of the instabilities.
The AC-SPH formulation presented here can therefore, at least qualitatively, 
correctly follow the time evolution of the cloud as in the other SPH schemes 
which have been recently proposed \citep{read10,hs10,mu11,sa12}.

\begin{figure*}
\centerline{
\includegraphics[width=15.2cm,height=8cm]{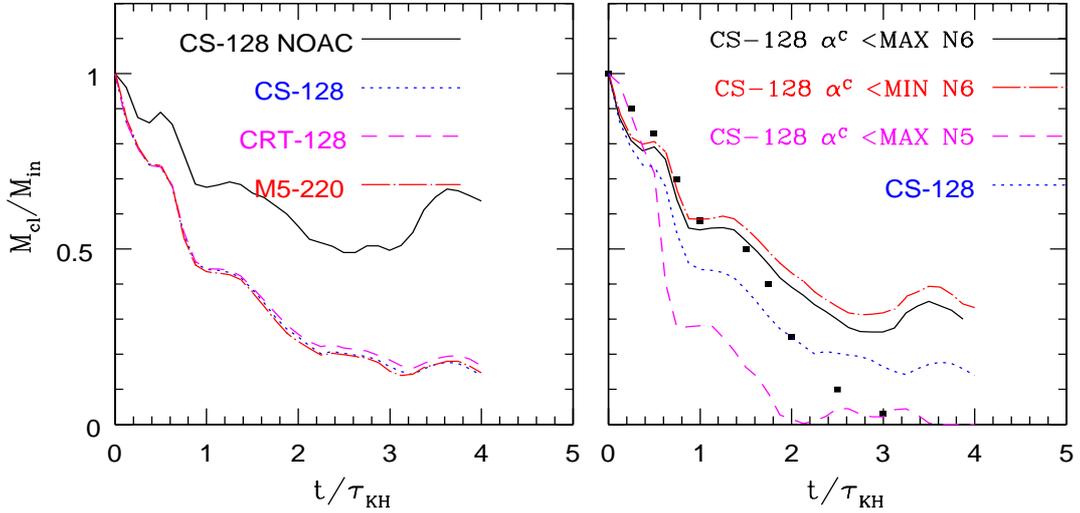}
}
\vspace{-0.5cm}
\caption{Mass loss of the cloud versus time for the blob test simulations. 
Particles are defined as cloud member if their temperatures and densities fulfill
 the conditions $T_i < 0.9 T_{ext} $ and  $\rho_i > 0.64 \rho_{cl}$.
Left panel: simulations have been performed incorporating the AC term into the 
SPH equations and using the kernels CS, CRT and $M_5$. For the CS kernel 
a simulation is also shown as reference without AC (NOAC, solid black line). 
Right panel: for the CS kernel simulations with additional constraints on
the  particle AC parameters $\alpha^C_i$  have been performed using $N=10^6$ (N6) 
and $N=10^5$ (N5) particles, see text for more details. 
Black squares have been extracted from Fig. 10 of \cite{sp11} and indicate 
the behavior of the cloud mass versus time in a similar test 
performed using the moving-mesh code AREPO  and with a number of 
$(64\times64\times128)$ resolution elements.
}
\label{fig:mloss}
\end{figure*}

To quantify the code performances in a more quantitative way , Fig.~\ref{fig:mloss} 
(left panel) shows  the mass loss of the cloud as a function of time for 
the different runs. 
We follow previous definitions \citep{ag07} and a gas particle is defined 
to  be still  member of the cloud if at the considered epoch its gas and 
temperature fulfill the conditions $\rho > 0.64 \rho_{cl}$ and  $T < 0.9 T_{ext}$.
The standard SPH results show a cloud that is not disrupted and its mass, 
after an initial transient, stays constant at about half of the original value. 
A striking result is instead given by the runs employing the AC term. 
For these simulations there is now a large mass loss rate occurring at 
early times, followed by complete cloud disruption at $t\simgt 3 \tau_{KH}$.

Note that the mass loss rate of the the AC-SPH runs 
does not depend in a significant way on the choice of the kernel. This suggests
that the cloud disruption is driven by large-scale instabilities and is
relatively insensitive to small-scale perturbations.
Given the similarities displayed in Fig.~\ref{fig:mloss}, left panel, by the 
mass loss rates of the AC-SPH simulations employing different kernels, 
when referring to the left panel we will adopt the term 
mass loss rate to generically indicate the behavior of these curves.

Clearly, in the test considered here the new AC scheme is now capable of 
properly removing the surface effects, which are present across the contact 
discontinuity in the standard SPH version, that artificially suppresses the growth
 of
hydrodynamic instabilities. 
However, a comparison of the mass loss rate with the corresponding one 
produced using the new moving-mesh code AREPO in simulations of
equivalent resolution \cite[Fig.10]{sp11}, shows that for the AC runs 
presented here the mass depletion of the cloud occurs much faster.
This discrepancy suggests that the processes of heat diffusion which 
in the adopted numerical scheme are mediated via the AC parameter $\alpha^C$, 
should be somehow constrained by a physically motivated mechanism which has 
not been previously considered in the discussion of Sect. \ref{acsph.sec}.
This mechanism should be introduced with the purpose of avoiding a
heat transfer mechanism, as governed in the code by Eq.~(\ref{duc.eq}), 
that is  overly diffusive.

In order to evaluate the relative effectiveness of heat and momentum
transport, in the theory of heat transfer the Prandtl number $Pr$ is defined 
 as the ratio between the kinematic viscosity $\nu$ and
the thermal diffusion coefficient $D$:  $Pr={\nu }/{D}$ \citep{bl06}.
For gases, the transport coefficients for the transport of heat and momentum are 
nearly equal and the Prandtl number is of order unity, 
with $Pr=\frac{2}{3}$ when $\gamma=5/3$ \citep{bl06}. This suggests that a 
constraint on the particle AC parameter $\alpha^C_i$  can be obtained by setting

\be
 Pr \simeq \frac{\nu_{AV}}{D_{AC}}\sim \frac{1}{5}
 \frac{\alpha^{AV}_{i} v^{AV}_{ij} }
  {\alpha^{AC}_{i} v^{AC}_{ij} } \simgt \frac {2}{3}~,
 \label{prac.eq}
\ee

where the equivalent kinematic viscosity coefficient $\nu_{AV}$ due to AV 
 is given approximately by 
$\nu^{AV}_i \sim \frac{1}{10} {\alpha^{AV}_{i} v^{AV}_{ij} r_{ij}}$
\citep{lp10}, and the numerical heat diffusion  coefficient $D^{AC}_i$ 
has been estimated  according to Eq.~(\ref{dac.eq}).
Note that the reverse of Eq.~(\ref{prac.eq}) is relatively unimportant, 
since the source term ${S^C}_i$ is driven by a second order derivative, 
whilst for AV is a first order derivative which determines ${S}_i$.

It can be seen that from Eq.~(\ref{prac.eq}), for the AC signal velocity adopted 
here, the constraint on the numerical heat diffusion becomes important in 
presence of supersonic flows. Moreover, the condition $Pr \sim O(1)$ is valid
only for gases, where the transport of momentum and energy occurs simultaneously. 
For liquids, according to the dominant mechanism of heat conduction,
the value of the Prandtl number can vary  by several orders of 
magnitude among different substances \citep{di05}. 
 The condition (\ref{prac.eq}) should, therefore, be considered problem dependent.

A rigorous procedure to derive the upper limits on the set 
of parameters $\left\{\alpha^{C}_{i} \right\}$ at any given timestep 
would require to first define 
for the particle $i$ the 
numerical kinematic viscosity and heat diffusion coefficients

\be
\left\{
 \begin{array}{ l l l }
< \nu^{AV}_i>  & = & | \left (\frac {d \vec v_i}{dt}\right )_{AV} | \, 
\big {/}  \,  |\nabla^2 \vec v_i +2 \vec \nabla \; (\vec \nabla \cdot 
\vec v_i) |  \\
<D^{AC}_i>  &  = & \left ( \frac {d u_i}{dt} \right)_{AC} \big {/} \nabla^2 u_i~,
 \end{array}
\right.
\label{numcof.eq}
\ee

where the bulk viscosity $\zeta$ is $5/3$ of the shear viscosity $\eta\equiv 
\nu \rho $ \citep{lp10}  and it is understood that the SPH expressions 
should be used for the operators at the denominators. 

The upper limits on the parameters $\left\{\alpha^{C}_{i} \right\}$  are then 
given by the conditions 
\be
<D^{AC}_i>\, \leq \frac{3}{2} < \nu^{AV}_i>~,
 \label{dclim.eq}
\ee
which must be simultaneously satisfied by all the parameters. 
In place of the procedure just described, we adopt here a simpler approach and 
using Eq.~(\ref{prac.eq}) we estimate the upper limits on 
$\left\{\alpha^{C}_{i} \right\}$ as

\be
 \alpha^{AC}_i \simlt \frac{3}{10} \alpha^{AV}_{i} MAX_{j<} 
\left\{ |v^{AV}_{ij}| /|v^{AC}_{ij}| \right\} ~,  
 \label{aclim.eq}
\ee

where the notation $j<$ means that the maximum is taken form all the 
neighbors $j$ that satisfy the  condition $ \vec v_{ij} \cdot \vec r_{ij}<0$.
This is to be consistent with the definition of AV, for which 
  $\Pi_{ij}$ is non-zero only for approaching particles. 
Clearly, this definition does not guarantee an accurate estimate of the 
constraint on the $\alpha^{C}_i$  parameter, nonetheless the use of the 
maximum among all the approaching pairs should provide a floor value 
for the effective constraint.

For the CS kernel, we then performed a simulation identical to the one
previously discussed, but now imposing the additional constraint 
 (\ref{aclim.eq}) on the $\alpha^{C}$  parameters.
This simulation is labeled as MAX N6 and the corresponding 
mass loss curve is displayed in the right-hand panel of Fig.~\ref{fig:mloss}. 
In order to assess resolution effects, we ran a mirror simulation now using 
$N=10^5$ particles (MIN N5). Moreover, the uncertainties associated with 
the use of Eq.~(\ref{aclim.eq}) can be estimated  by looking at the mass loss 
rate of a simulation (MIN N6) where instead  of  Eq.~(\ref{aclim.eq})  
the same equation was used but the constraint  is derived using the 
$MIN$ operator instead of the $MAX$ one.

For comparative purposes, in the right-hand panel of Fig.~\ref{fig:mloss}
the mass loss of the cloud (black squares), as found in 
a similar blob test performed by \cite{sp11} using the new moving-mesh code 
AREPO, was also inserted. The number of resolution elements is $64\times64\times128$, so that
the resolution is approximately equivalent to that of the tests displayed here.

A striking result seen from the behavior of the mass loss rates 
is that, introducing  the constraint (\ref{aclim.eq})  on the
numerical heat transfer in the simulations, the mass depletion of the cloud  is now in better 
agreement with previous results  and in particular with the cloud mass evolution
 produced in the blob test by the Voronoi-based code AREPO. 

Uncertainties associated with the procedure adopted to estimate the constraint
on the $\alpha^{C}$  parameters are of limited impact, as can be assessed by the 
differences between the mass loss rates exhibited by the run MAX N6 and MIN N6.
Moreover, for the run MAX N5 the mass loss rate is much stronger than in the case 
MAX N6. This is indicative of numerical diffusion effects which dominate the
blob evolution. Clearly, a numerical simulation with $N=10^7$ particles 
should be required to clarify this issue; however the agreement 
 with previous results of the mass loss for the run MAX N6 suggests that 
convergence is already being achieved using $N=10^6$ particles. 

Finally, both the runs MAX N6 and MIN N6 exhibit at $t \simgt 3 \tau_{KH}$ 
a remnant cloud mass which is 
of the order of $\sim 20-30\%$ of the initial mass. 
Although such a 
result can still be due to the use of a constraint on the numerical heat
diffusion which still needs to be refined, we note that such remaining masses
  are equally present in SPH blob tests which have adopted, 
with the purpose of avoiding the problems of standard SPH,
completely different approaches \citep{hs10,mu11} 

We, therefore, suggest that this underestimate in the stripping rate of 
the blob mass is not due to the AC scheme, but depends rather on the 
SPH formulation and is associated with the intrinsic errors in gradient estimates.
These errors, in turn, lead to the suppression of the small-scale instabilities.
Such a issue will be discussed further in detail in Sect. \ref{RT.sec}
dedicated to the Rayleigh-Taylor  instability.

To summarize,  SPH simulations which incorporate the AC  scheme can 
successfully be used to accurately model the blob evolution. However, 
the results of the tests indicate that it is necessary to properly constrain
the AC parameter $\alpha^{C}$ in order to avoid unphysical heat diffusion
when strong supersonic flows are present.

\section{Gravity tests}
\label{hygrav.sec}

To validate the performances of the AC signal velocity (\ref{vsgv.eq}), several 
hydrodynamical test problems were investigated in which the self-gravity must be 
necessarily taken into account to properly model the system evolution.
We first consider the 3D collapse of a cold gas sphere initially at rest, 
this test being customarily used in SPH to judge the code capability when strong 
shocks are present, such as those occurring during the formation of 
self-gravitating structures. We then examine a 2D version of the classic 
Rayleigh-Taylor instability  test and finally the new code is used to assess 
the behavior of cluster entropies in cosmological structure formation.

\subsection{Cold gas sphere}
\label{cold.sec}

\begin{figure*}
\centerline{
\includegraphics[width=15.2cm,height=15.2cm]{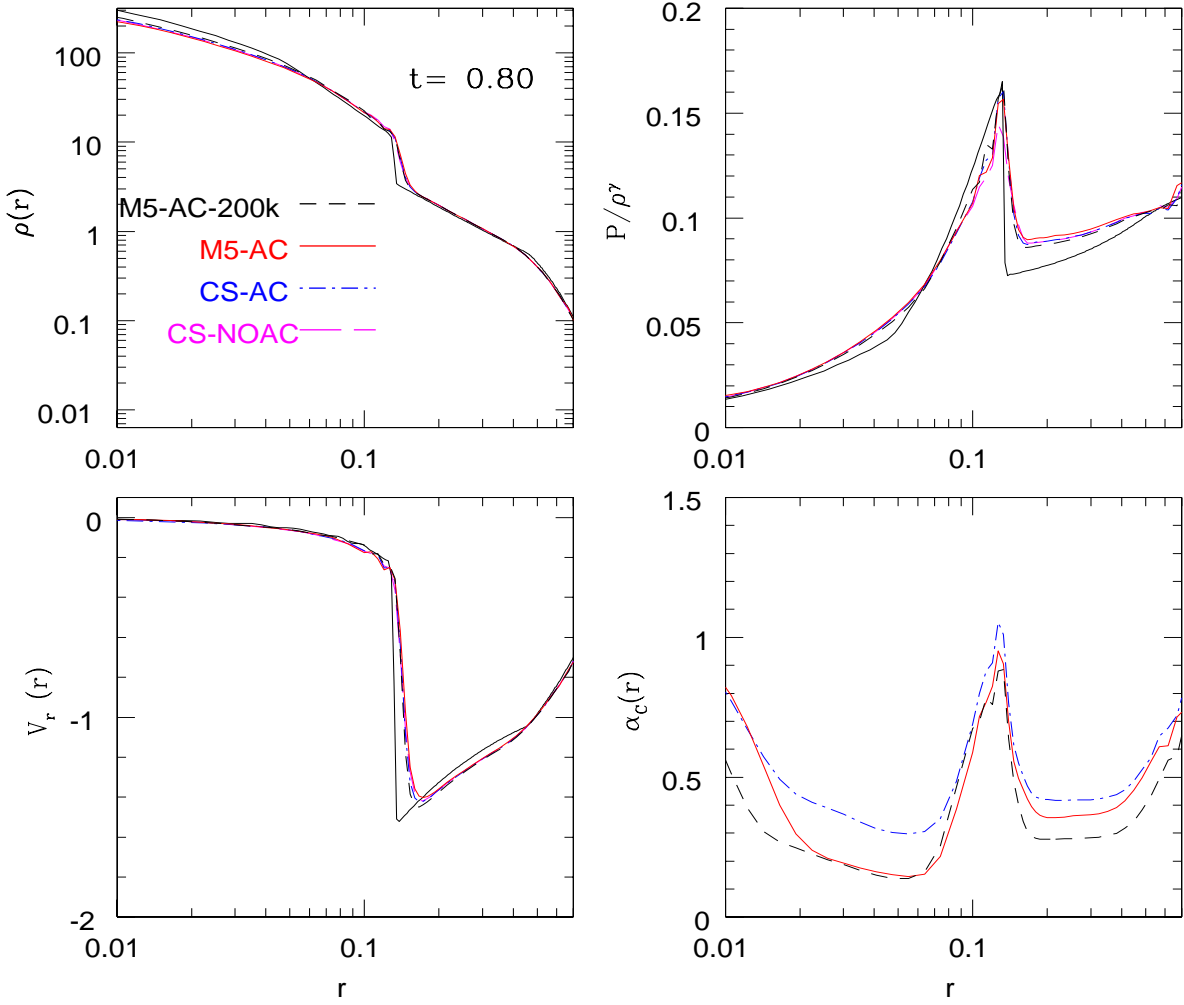}
}
\vspace{-1cm}
\caption{Radially averaged profiles at $t=0.8$ of the Evrard collapse test.
Clockwise from the top left panel:  profiles of
density, entropy, artificial conductivity parameter $\alpha^C$ and radial 
velocity. Curves with different line styles and colors refer to SPH runs
performed using different kernels and with (AC) or without (NOAC) the AC term
in the SPH energy equation.
The black solid lines indicate the 1D PPM reference solution of \cite{st93}.}
\label{fig:coldb}
\end{figure*}

A standard hydrodynamical test for SPH codes  in which gasdynamics is modeled 
including self-gravity is the 3D collapse of a cold gas sphere, 
also commonly recognized as the 'Evrard'  collapse test \citep{ev88}.
The test follows in time the adiabatic collapse of a initially cold gas sphere  
and has been widely used by many authors 
\citep[V11]{hk89,st93,wa04,sp05,we09,sp10b,hs10} 
as a standard test for SPH codes.

The gas cloud is spherically symmetric and initially at rest, with mass $M$, 
radius $R$, and density profile
   \begin{equation}
   \rho(r)=\frac{M}{2\pi R^2}\frac {1}{r}~.
  \label{rhocl.eq}
   \end{equation}

The gas obeys the ideal gas equation of state with $\gamma=5/3$ and the
thermal energy per unit mass is initially set to $u=0.05 GM/R$. 
The SPH simulations are performed using units for which $G=M=R=1$
and the chosen time unit is the cloud free-fall timescale 
$t_{ff}=(\pi^2/8)\sqrt{R^3/GM}=\pi^2/8$.

With these initial conditions, the pressure support of the gas sphere is 
negligible and the cloud begins to collapse until a bounce occurs in the 
core with a subsequent shock wave propagating outward.
Most of the kinetic energy is converted into heat at the epoch
of maximum compression of the gas, which occurs at $t\sim1.1$.
The initial conditions setup is realized   by stretching the radial
coordinates of a glass-like uniform distribution of $N=88,000$ equal 
mass particles located  within a sphere of unit radius, so as to generate the 
density profile of Eq. (\ref{rhocl.eq}). 

To construct the set of SPH simulations, four tests have been run using the 
kernels CS and $M_5$. The number of neighbors is set to $100$ for 
the $M_5$ runs and $50$ for the CS ones.
To assess convergence properties in the radial profiles of the considered 
 hydrodynamic variables, the $M_5$ run has been replicated 
using $200,000$ particles 
($M_5-200k$). All of these runs were performed incorporating in the 
SPH equations the AC term (\ref{duc.eq}), for the CS kernel a reference run 
with the AC term  disabled has been considered (CS-NOAC).
The gravitational softening length is taken as $\varepsilon_g=0.02$. 

For each test case, we perform simulations with different settings of the AV 
parameter $l_d$. However, introducing a time-dependent AV scheme in SPH 
 reduces numerical viscosity effects and in  particular, for the test investigated 
here, produces  a radial entropy profile  at the shock front in better agreement 
with the 1D PPM reference solution.
For this reason, the results for different test cases are presented for 
those runs where a standard AV formulation has been used ( runs $AV_0$ of V11). 
This is done in order to highlight differences in the simulation results 
due to the use of different kernels and of incorporating the AC term in the SPH
thermal energy equation.
This choice for the simulation parameters
 allows one to compare the test calculations performed here with previous
results presented in  Sect. 5.2  of V11.

The average radial profiles at $t=0.8$ of
 density, entropy, time-dependent AC parameter and radial velocity are displayed 
in Fig.~\ref{fig:coldb}. 
The black solid lines indicate the profiles of the 1D PPM calculation 
of \cite{st93}.
Broadly speaking, one expects to see some differences between the profiles of 
the NOAC simulation and those of the other runs. This should be valid in particular
for the entropy profiles. However, at the considered epoch,  Fig.~\ref{fig:coldb} 
shows that the entropy profile of the NOAC run is still very similar to the 
others. This occurs because at  $t=0.8$ the shock front propagation is still in 
the early phase and the smoothing in the thermal energy due the AC term 
is not yet significant.

The behavior of the $\alpha^C$ profiles is in accordance with what expected. 
In particular, the profiles exhibit a peak in correspondence of the 
shock front location, which occurs approximately at $r\sim0.18$, and a 
decay as one moves away from it. There is a dependence on the kernel shape
and a weak dependence on resolution. The former is interpreted as a 
consequence of an improved mixing capability due to the increased order of 
the kernel. 
Note that there is a rising in 
the profiles as one moves toward the sphere center, this occurs because of
the presence of a temperature gradient.
However, in Eq.~(\ref{duc.eq}) this radial dependence is  of no effect 
since behind the shock front the sphere quickly achieves a hydrostatic equilibrium 
and $v^{AC}_{ij}(grav)\simeq0$ for $r\simlt 0.1$.

The epoch examined in  Fig.~\ref{fig:coldb} has been chosen so as to compare the
simulation profiles with the corresponding ones presented in Fig.~40 of
\cite{sp10b} and realized using the new moving-mesh code AREPO.
In particular, a visual inspection shows that ahead of the shock front the 
accuracy of the entropy profiles shown in Fig.~\ref{fig:coldb} can be 
considered intermediate between the profile produced using the AREPO code 
and the one realized with the SPH code Gadget-2 ( right panels in Fig.~40 
of \cite{sp10b}, test cases `moving-mesh' and `SPH', respectively).
Note, however, that a strict comparison between the different simulation 
profiles is difficult because the number of resolution elements  
employed by \cite{sp10b}  in the cold gas sphere tests are lower by about 
$\sim 1/3$ than the one used here. In particular, the smaller shock broadening 
found here can be ascribed to a resolution effect.
\begin{figure*}
\centerline{
\includegraphics[width=16cm,height=8cm]{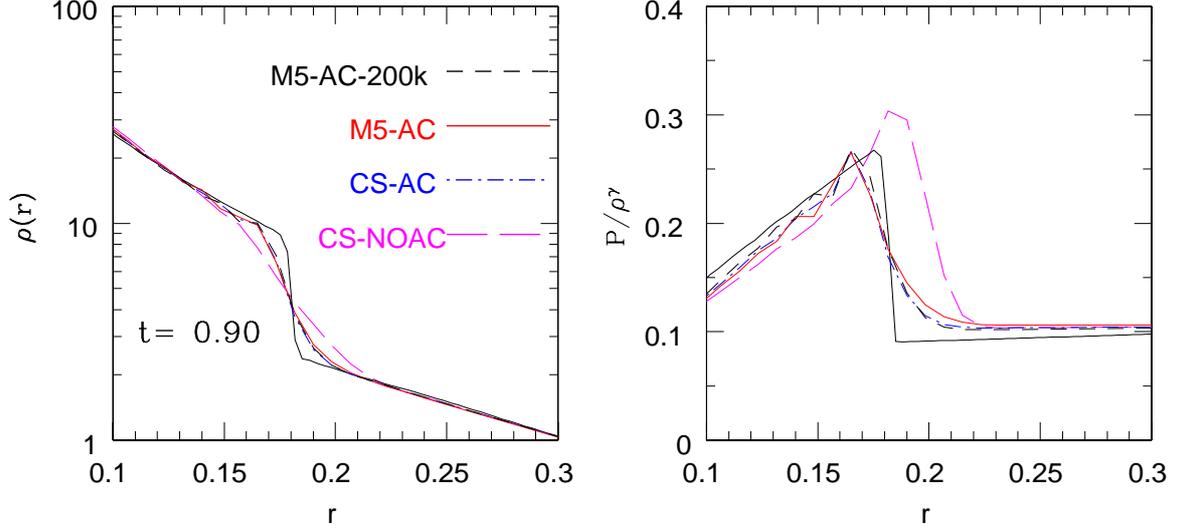}
}
\caption{As in Fig.~\ref{fig:coldb}, a closer view at $t=0.9$ of the    
radial profiles of density and entropy in the proximity of the shock front.}
\label{fig:coldc}
\end{figure*}

Finally, Fig.~\ref{fig:coldc} shows a closer view of the radial
density and entropy  profiles in the proximity of the shock front at $t=0.9$.
The NOAC simulation exhibits now the worst accord with the PPM reference solution,
while for the high-resolution run $M_5-200k$ the improvement in accuracy is 
minimal with respect to the other AC runs.
This suggests that to obtain a significant improvement in the  adherence of the 
simulation profiles to the PPM solution one must use a much higher number of 
simulation elements ( say $\simgt 10^6$, \cite{sp10b} ).

A noteworthy feature of the profiles of Fig.~\ref{fig:coldc} is that now the 
AC simulation profiles are in much better agreement with the PPM profiles. 
This is interpreted as a numerical effect in which the pre-shock entropy, 
which is generated by the AV implementation ahead of the shock front owing to 
the converging flow,  is now strongly reduced because the presence of the AC term
in the SPH energy equation now removes this excess of internal energy.
  This in turn implies, behind the shock front, a thermal energy profile  closer  
to the PPM solution profile.

Comparisons of the profiles of Fig.~\ref{fig:coldc} with the corresponding 
ones displayed in Fig.~4 of V11 show that the dependency of the simulation
profiles on the adopted AC scheme are stronger than the ones due to the
time-dependent AV formulation. In fact, these results are almost unaffected if
one performs SPH simulations using now a time-dependent AV scheme in place of 
the standard one.
 
Finally, two runs have been performed using the CRT and LIQ kernels 
and a number of neighbors set to  $N_{sph}=50$. The profiles of the two
simulations have  not been shown here to avoid overcrowding in the 
panels, however the 
performances of the CRT run are quite similar to those of the CS one, whilst
the profiles of the LIQ simulation exhibits very poor  consistency properties
with respect to the reference PPM solution profiles.

\subsection{2D Rayleigh-Taylor}
\label{RT.sec}

The Rayleigh-Taylor (RT) instability arises when 
a heavier fluid is placed on the top of a lighter fluid 
\citep{ch61} in presence of an external gravitational field.
  The fluids are in pressure equilibrium against the external 
field and in this
configuration the system is unstable in the presence of small perturbations
at the interface, the lighter fluid will then begin to rise and the denser 
one to fall. This process leads to the development of characteristic finger-like
structures before the fluids enter the non-linear phase where they  
mix together completely.  In order to evaluate the ability of the AC-SPH code 
to properly describe the evolution of RT instabilities,
 we consider  a 2D version of 
the test. The initial conditions are chosen to be similar to the numerical test 
implemented by \cite{ab11} to validate its new rpSPH formulation of SPH 
equations, so as to consistently compare the results.
 The computational domain consists in a 2D box with coordinates 
 $x\in \{ 0,1/2 \}$, $y\in \{ 0,1 \}$.  The boundaries are periodic in $x$ and 
reflecting in $y$. The density is $\rho_1=2$ at the top and $\rho_2=1$ at the 
bottom, with a density profile

\be
\rho(y) = \rho_2+\frac{(\rho_1-\rho_2)}{\left[ 1.+
\exp{\{-2(y-0.5)/\Delta_y\}}\right]}~,
\label{eq:rht1}
\ee

where $\Delta_y=0.05$. This ensures  a smoothing in density at the 
interface $y=0.5$
that allows a consistent numerical behavior \citep{rb10}.
This density profile is realized by constructing a HCP lattice of $N=620^2$ 
equal mass particles in which the spacing is varied according to the 
procedures described in Sect. \ref{inkh2d.sec} until the profile 
(\ref{eq:rht1}) is satisfied.

The pressure at the interface is set to $P_0=\rho_1/\gamma=10/7$, where 
$\gamma=1.4$ and it varies with $y$ according to $P(y) = P_0 -g \rho(y) (y -1/2)$, 
$g=1/2$ being the external acceleration, so that the system is initially 
in hydrostatic equilibrium.
For the particles that satisfy the condition $|y-0.5|<0.2$ a velocity
perturbation is applied in the $y$ direction given by

\begin{eqnarray}
v_y(x,y) & = &\delta v_y \{ 1+ \cos [ 8 \pi (x +1/4) ] \} \nonumber  \\
   & & ~~~~~\{ 1+ \cos [ 5\pi (y -1/2) ] \}/4~,
\label{eq:rht2}
 \end{eqnarray}

where $\delta v_y=0.1$.

We present results for the CS, CRT, $M_5$ and additionally, for reasons 
which will be discussed later,  we consider 
also the $M_6$ kernel \cite[sect. 2.3]{pr12}. 
 The number of neighbors for these runs is set to $N_{sph}=32,32,50,110$, 
respectively.
As in the other test cases considered here, for comparative purposes, we ran a 
simulation (CS-NOAC) in which the AC term in the thermal energy equation has 
been disabled. The following settings have been used for the AV parameters:
 $\{ \alpha_{min}, \alpha_{max}, l_d\}=\{0.1,1.5,0.2\}$. 

\begin{figure*}
\centerline{
\includegraphics[width=18.2cm,height=8.5cm]{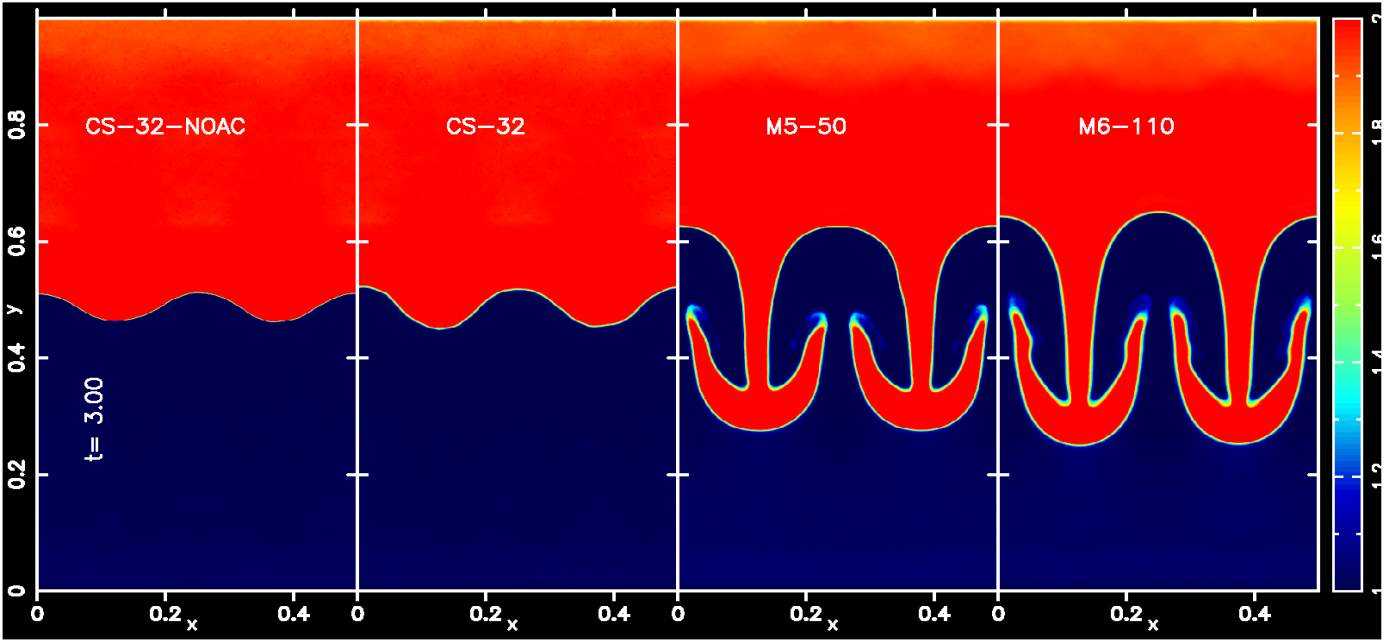}
}
\caption{ Density maps of the 2D Rayleigh-Taylor tests are shown at time  $t=3$ 
for standard SPH (NOAC) and AC-SPH simulations performed using $N=620^2$ particles 
  with different kernels and neighbor numbers.
The density distributions can be compared directly with the map shown 
in Fig. 4 of \cite{ab11}, top right panel, as the adopted 
initial conditions for the tests are the same.} 
\label{fig:rtmap}
\end{figure*}
In Fig.~\ref{fig:rtmap}, at the time $t=0.3$, we present the 2D density maps 
of the different RT tests.
As expected, standard SPH is unable to correctly capture the development of 
the RT instability, as indicated by the leftmost panel of Fig.~\ref{fig:rtmap}
 (CS kernel, NOAC run). However, from this viewpoint the improvement is 
minimal even for the corresponding AC version of the considered run (CS-32). 
On the contrary, a significant improvement is obtained if one uses the $M_5$ kernel
or the next in order $M_6$, which is a quintic polynomial \citep{pr12}. In fact, 
in the rightmost panel of Fig.~\ref{fig:rtmap} ($M_6$ run), the center of mass of 
the RT spikes appears located at a lower position than in the $M_5$ run and 
therefore the convergence might still not be achieved even for the $M_6$ run.

This strong dependence of the code performances on the kernel order 
demonstrates that, for the RT test, accuracy in pressure forces 
is a fundamental issue. These results are consistent with those of 
sect. \ref{kh2d.sec} and indicate that the poor performances of SPH when 
handling KH or RT instabilities, in particular for very subsonic flows, 
are mainly due to the leading errors in the momentum equation.
These errors are reduced as the order of the kernel is increased, 
therefore implying  an improved accuracy in pressure force estimates and
thus  a lower velocity noise. This, in turn, implies a better capability 
to capture the growth of the instability.

Similar results have recently been obtained by \cite{na11}, who ran a suite of
2D KH simulations using  carefully crafted initial conditions with the 
goal of assessing different hydrodynamic code capabilities. 
The authors conclude that for 
SPH the code performances are strongly related to the order of the kernel 
employed in the simulation and thus to the accuracy of the gradient estimates.

Moreover, the simulation behavior of Fig.~\ref{fig:rtmap}
is consistent with the findings of \cite{ga12}.
The authors present a new formulation of SPH in which gradients are estimated
using a tensorial approach which conserves both linear and angular momentum.
The authors demonstrate that using this higher order gradient estimator 
the interpolation of physical quantities is significantly improved with 
respect to  the standard method. 
In particular, several tests showed that the code is now able to successfully 
capture the development of KH and RT instabilities.

To summarize, the results of this section demonstrate that 
 SPH performances to model hydrodynamic subsonic instabilities 
depend critically  on the accuracy of the gradient estimator.
The formulation proposed by \cite{ga12} looks particularly promising in this
aspect to overcome the present SPH difficulties.

Finally, it must be stressed that the simulation results of Fig.~\ref{fig:rtmap}
have been obtained setting $\delta v_y=0.1$. If the same test had been
performed using  $\delta v_y=0.01$, as in the RT test of the new moving-mesh
code  AREPO \cite[Sect. 8.8]{sp10b}, the velocity noise would have suppressed
the growth of the instability even for the $M_6$ run. Note, that in their
RT test, \cite{ga12} were able to recover  the growth of the instability 
using the same amplitude of the velocity perturbation, $\delta v_y=0.01$.

\subsection{Cluster comparison}
\label{cluster.sec}

In this section, a set of cosmological cluster simulations is constructed 
using the standard SPH formulation as well as the new AC-SPH version. 
We only consider non-radiative, or 'adiabatic' simulations, in which the 
hydrodynamic is modeled according to the formulation presented in Sect. 
\ref{hymeth.sec}. Previous investigations \citep{fr99,wa04,wa08,mi09,sp10b} 
showed that for the same initial conditions there are systematic differences 
between results extracted from cluster simulations produced using standard 
SPH and Eulerian AMR codes. Specifically, the level of central entropy 
is found to be lower in SPH simulations than in the corresponding AMR runs. 
The latter simulations are characterized by the presence of a flat 
entropy core, whereas the radial entropy profile produced by SPH runs 
increases steadily with 
radius. As already outlined in the Introduction, the origin of this discrepancy 
has been the subject  of an intense debate \citep{ag07,mi09,read10} which has 
given as main explanation the different 
degrees of numerical mixing present in the two codes.

To assess the capability of the AC-SPH code  in solving this issue, we then run 
several cluster simulations. A comprehensive study of the differences between 
the hydrodynamic variables of cluster simulated using the standard and the AC 
version  of the SPH code will be presented in a forthcoming paper. Here, 
we just show the final radial entropy profile of the simulated clusters 
as the main variable that can be used to test the effectiveness of the
AC-SPH formulation for solving the entropy problem in cluster cores.

A detailed description of the procedures used to construct the simulated 
cluster sample is given in sect. 2 of V11, we provide a brief
summary here. The simulations were carried out assuming a spatially flat 
${\Lambda}$CDM  model, with 
matter density parameter $\Omega_\mathrm{m}=0.3$, vacuum energy density 
$\Omega_\mathrm{\Lambda}=0.7$, baryonic density $\Omega_\mathrm{b}=0.0486$
 and $h=H/100 km sec^{-1} Mpc^{-1}=0.7$. The scale-invariant power spectrum 
is normalized to $\sigma_\mathrm{8}=0.9$ on an $8 \, h^{-1}$ Mpc scale at 
the present age $t_0$.
An N-body cosmological simulation involving only gravity was first run 
with a comoving box of size $L_2=200h^{-1}$Mpc
starting from an initial time $t_{in}$ down to the final epoch $t_0$.
At this  epoch clusters of galaxies are identified in the simulation box 
as groups of particles that are associated with overdensities 
 approximately in excess of $\sim200 \Omega_\mathrm{m}^{-0.6}$.
Several of these clusters are then resimulated individually using the 
AC-SPH code here described, coupled with a treecode gravity solver.

Initial conditions for a specified cluster are generated as follows:
a spherical region with origin located at the cluster center is
populated with a high-resolution (HR) grid, the radius of the HR sphere 
is such that it contains all of the original particles identified at 
$t=t_0$ as cluster members. 
A gas and a dark matter particle is associated to each grid node,  
whose positions are perturbed according to the 
random realization of the original cosmological simulations and only 
those particles which are located within the HR sphere are kept for 
the hydro run. The HR particles are surrounded by a low-resolution shell
of dark matter particles, extracted from a grid with spacing twice that of the
 HR grid, the shell is introduced with the purpose of mimicking the effects of
tidal forces. The grid spacing is chosen such that at the end of the procedure
a cluster is simulated with $N_{gas}\sim 220,000$ gas particles and $N_{dm} \sim
N_{gas}$ dark matter particles in the inner HR sphere,  whereas 
$N^{ext}_{dm}\sim N_{dm}$ are used in the low-resolution shell.
Particle masses are assigned according to the values of
$\Omega_\mathrm{m}$ and $\Omega_\mathrm{b}$.

The clusters selected to be re-simulated hydrodynamically were chosen 
with the following criterion. Originally (V11), the procedure previously 
described was repeated two more times in order to
 construct two new cluster samples  extracted from cosmological simulations with 
box sizes $L_4=2L_2$ and $L_8=2L_4$. The three cluster samples were then 
combined to construct a final sample $S_{all}$ of $\sim 160$ clusters
covering nearly a decade in cluster masses. Cluster members of sample $S_{all}$ 
were then ranked according to their dynamical state, as measured by an appropriate 
statistical indicator. Two sub-samples of four test clusters each, denoted 
respectively by $Q$ and $P$, were then constructed by extracting from sample 
$S_{all}$  those clusters with membership criterion  
 for sample $Q$ ($P$) of being in a fully relaxed (perturbed) state. 
These two sub-samples were then used to investigate 
the effects of AV in hydrodynamic SPH simulations of galaxy clusters (V11).
Here the four test clusters of sub-sample $Q$ were chosen to investigate the 
performances of the AC-SPH code. This choice being motivated by the need  to
proper disentangle  
the amount of entropy mixing produced during merger processes from that specific 
to the numerical method in the final cluster entropy. 
The clusters of sub-sample $Q$ were carefully selected on the basis of their 
dynamical state and are among the most relaxed of sample $S_{all}$. Therefore,
for these clusters, differences in the final radial entropy profile between 
standard and AC runs can be safely ascribed to the new AC term implemented 
in the code. The AV simulation parameters of the simulations are $\{ \alpha_{min}, \alpha_{max}, l_d\}=\{0.1,1.5,0.2\}$. 
\begin{figure}
\hspace{-0.7cm}
%\vspace{0.5cm}
\includegraphics[width=10.2cm]{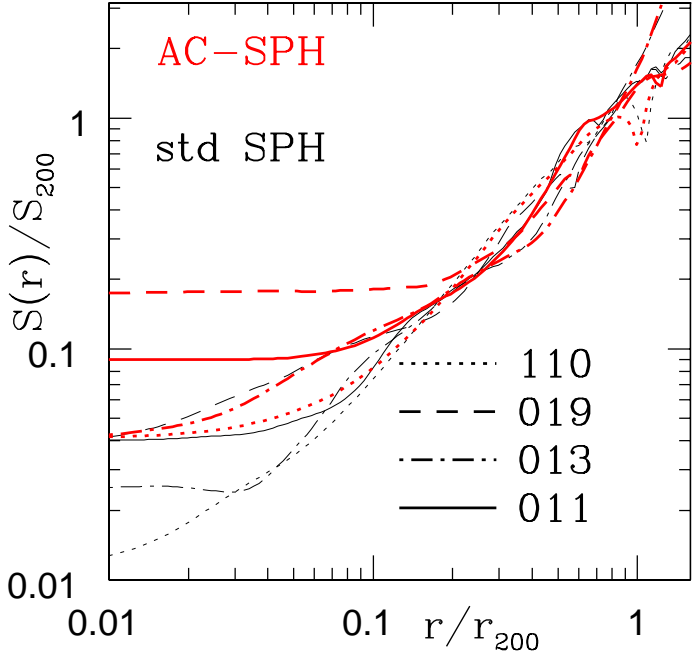}
\caption{Final radial entropy profiles as a function of $r/r_{200}$ 
for the four relaxed test clusters. The gas entropy
$S(r)=k_B T(r)/\mu m_p \rho_g^{2/3}$ is plotted  in units of $S_{200}$ 
Different line styles refer to different clusters, the numbers indicating
the cluster membership in the cosmological ensemble from which they 
have been extracted. Thin (black) lines are for the profiles 
of runs performed using the standard SPH formulation, while thick (red) 
lines refer to the AC-SPH runs incorporating the AC term.}
\label{fig:sprof}
\end{figure}

The final entropy profiles for the four test clusters are displayed in 
Fig. \ref{fig:sprof}, where the cluster entropies have been normalized to

   \begin{equation}
S_{200}=\frac{1}{2}\left [\frac{2 \pi}{15} \frac {G^2 M_{200}}{f_bH}\right]^{2/3}~,
  \label{entr.eq}
   \end{equation}

 where $f_b=\Omega_\mathrm{b}/\Omega_\mathrm{m}$ is the global baryon fraction
and $ M_{\Delta}= (4 \pi/3) \, \Delta\, \rho_\mathrm{c} \, r_{\Delta}^3 $
 denotes the mass contained in a sphere of radius
$ r_{\Delta}$ with mean density $\Delta$ times the critical density 
$\rho_\mathrm{c}=3H^2/8\pi G$. As is customary, the cluster mass 
is defined setting $\Delta=200$ and $r_{200}$ denotes the cluster radius.

The results of Fig. \ref{fig:sprof} indicate that, although there is a
certain degree of scatter between the entropy profiles of individual clusters, 
nonetheless all of the AC-SPH runs consistently exhibit  
 much shallower entropy profiles and higher core
entropies than in the standard SPH runs at $r/r_{200}\simlt 0.1 $.
For a given cluster, the difference in the levels of central entropies produced 
by the two codes is about of a  factor $\sim 4$, and now the values of central 
entropies 
are comparable with those produced using AMR codes in cosmological cluster
simulations \citep{vog05} and in idealized cluster binary mergers \citep{mi09}.
It should, however, be stressed that in Eulerian hydrodynamics it is the
numerical scheme that forces the fluid to be mixed below the minimum cell size.
This suggests that AMR simulations of galxy clusters might overestimate the
correct level of core entropy because of fluid mixing \citep{sp10b}.
To summarize, the development of entropy cores in the AC-SPH runs is clearly a
 consequence 
of the heat diffusion term (\ref{duc.eq}), now present in the energy equation.
This term acts to redistribute the internal energy produced in the shocks 
during dissipative processes, so that the subsequent entropy mixing leads,
in the end, to higher levels of core entropies. 

Similar results were obtained by \cite{wa08}, 
who similarly added a heat diffusion term to the SPH energy equation, 
but using the prescription (\ref{ducw.eq})~instead of the Eq. (\ref{duc.eq}) 
 adopted here. According to \cite{wa08}, the value of the coefficient $C$ tends to 
be problem dependent and for the  cluster simulations the best agreement 
with the entropy profiles extracted from the mesh code runs is obtained 
setting $C\simeq 0.1-1$.  Here, the time-dependent formulation
(\ref{alfac.eq}) presents the advantage of minimizing thermal diffusion 
away from shocks, as demonstrated in the cold gas sphere test by 
Fig.~\ref{fig:coldb}, where the radial profile of the AC parameter 
$\alpha^{AC}$ is peaked at the shock location.

The simulations presented in this section have been realized using adiabatic 
gas physics. However, a realistic modeling of the intra-cluster medium physics
must incorporate the possibility that the gas cools radiatively. Previous
simulations (V11) showed that the presence of cooling leads to the development
in the inner cluster regions of  dense compact cool gas cores 
and subsequently of high levels of turbulence, the latter being produced 
by the  hydrodynamical instabilities generated 
by the interaction of the compact cool cores with the ambient medium.

How this scenario is affected by incorporating into the SPH equations a 
numerical heat diffusion term is an issue which can be properly addressed only
by resorting to numerical simulations.
Here, we outline that two competing effects are expected to influence the final
level of turbulence in a cluster core: the improved capability of the code to
describe the development of hydrodynamical instabilities, and thus an increase
in the amount of turbulence, and the reduced availability of cool gas because 
of the presence of material of higher entropy which has been brought in the
inner core due to the enhanced fluid-mixing properties of the code.

Finally, a resolution study was already performed in V11, indicating 
for the simulation resolution employed here a substantial stability in the 
entropy profiles of the simulations.
Although the simulations of V11 do not incorporate the AC term, we do 
not expect strong variations in the resolution dependence of the profiles, 
given the similarity of the level of core entropies with those found
using mesh-based codes.

\section{Summary and conclusions}
\label{conc.sec}
In this paper, we have presented an SPH numerical scheme which incorporates 
an artificial conductivity term and uses an appropriate signal velocity for
simulations including gravity. The AC formulation has been introduced by 
 \cite{pr08} as a solution to the problems encountered by standard SPH to
correctly follow the development of KH instabilities, due to the 
inconsistencies of the standard formulation in the description of density at 
contact discontinuity. 
Here, a suite of hydrodynamic test problems is investigated with the purpose of
validating the new AC-SPH code and to assess its performances when using the
specifically adopted signal velocity.

The results of the KH instability test are presented in Sect. \ref{kh2d.sec},
in which the code capabilities have been tested by considering SPH simulations
of the KH test performed using a large variety of initial condition set-up 
and SPH kernels. The set of initial conditions has been chosen as in 
\cite{vrd10},
so as to consistently compare the results. These are in accordance with the  
corresponding ones of \cite{vrd10} and indicate that, for the version of the
KH test analyzed here, the AC implementation is important for the 
long-term behavior of the simulation. 

Moreover, the results of the KH test indicate \citep{vrd10,read10,na11}
 that the poor performances of standard SPH to
properly treat KH instabilities can be explained in terms of two distinct 
effects.
The first is a general problem of consistency, which for the 
problem under consideration requires that smooth interfaces should 
be present at contact
discontinuities, in order to obtain numerical convergence.
The second effect which in standard SPH suppresses the growth of KH instabilities
is the leading error in the momentum equation, due to incomplete kernel 
sampling \citep{read10} and quantified by the norm $\vec {E}^{0}$ 
defined by Eq. (\ref{en0.eq}).

To circumvent this problem several proposals have been made 
\citep{vrd10,read10} in which the standard SPH cubic spline $M_4$ kernel is
replaced by the new LIQ or CRT kernels with steeper central profiles.
These kernels present the advantage of being stable against particle clumping, 
so that the number of neighbors can be safely increased in order to reduce the 
error $\vec {E}^{0}$.

In this paper, we consider the possibility of reducing sampling errors by 
considering higher-order $B-$spline kernels, specifically the $M_5$ or quartic
spline kernel. 
A striking result of the KH runs presented in sect.  \ref{kh2d.sec} is that, 
for a 
given value of the ratio $\eta$ between the smoothing length and the mean 
interparticle spacing, simulations of the KH test performed using 
the $M_5$ kernel have amplitudes  of the  $\vec {E}^{0}$ error substantially 
smaller  and in line with the behavior of the same quantity for simulations
employing the LIQ or CRT kernels.
A linear stability analysis reveals that this result follows owing to the very
good stability properties of the $M_5$ kernel, in fact the analysis suggests 
that the clumping instability is absent for values of 
$\eta$ up to $\eta\simlt2$, 
which in 3D corresponds to $N_{sph}\sim520$ neighbors.

These findings are consistent with the recent results  of \cite{de12} who
proposed to adopt, with the specific purpose of avoiding the clumping 
instability, the \cite{we95} functions as a new class of kernels. 
The authors showed that in terms of stability and accuracy properties, 
the quartic spline 
performs extremely well when compared with the proposed Wedland functions.
These results are strictly connected to the properties of the kernel 
Fourier transform, which according to the authors must be non-negative
to avoid particle clumping, and are consistent with the findings of
sect. \ref{khstab.sec} since increasing the kernel order both the $B-$spline
and the Wedland functions approach the Gaussian.

We therefore propose, as a compromise between the need of reducing sampling
errors while keeping the computational cost at a minimum, the use of the 
$M_5$ kernel with  neighbor number in the range $N_{sph}\sim60-120$ as
the standard combination which guarantees sufficient accuracy in many 
SPH simulations of astrophysical problems.  
Moreover, the results of the Sod shock tube test of sect. \ref{sod.sec}
demonstrate that in order to obtain simulation profiles in accordance with the
analytic solution, for simulations employing kernels with a modified shape 
 the use of a much larger number of neighbors than  in the
case of the $M_5$ runs is necessary.

The results of the gravity tests show that the adoption of the AC-SPH scheme 
significantly reduces, at the level tested in this paper, the differences seen 
in the hydrodynamics between standard SPH and grid-based simulations of 
self-gravitating structures.
For the cold gas sphere, the entropy profile is in better agreement with the 
PPM reference solution and for the cosmological cluster simulations, a key 
result is the final level of core entropies,  which are consistent with 
those  of the central entropies produced  using AMR codes.
Thus, it appears that in hydrodynamic simulations where self-gravity 
is important the AC term, accompanied by the proposed signal velocity, 
plays a key role as a mechanism of redistributing thermal energy  and
hence as a source of entropy mixing.  

To summarize, results extracted from simulations of hydrodynamic tests where
self-gravity dominates are in much better agreement with the corresponding ones 
obtained using mesh-based codes. The results then demonstrate the capability 
of the implemented AC-SPH scheme to properly follow the formation of 
cosmic structures. 

It is worth noting that the artificial heat conduction term was 
originally proposed by \cite{pr08} with the purpose of avoiding the 
inconsistencies encountered by standard SPH in presence of density steps at
contact discontinuities. A complementary view has been proposed by 
\cite{wa08}, who introduced the same term, albeit in a different numerical
formulation, with the aim of modeling the level of diffusion due to turbulence.
 The two interpretations are not mutually inconsistent, however the results of
the self-gravity  tests presented in this paper support the view of a heat
diffusion term which in SPH is capable of mimicking the diffusion due to
turbulence. In a similar fashion, \cite{vi07}  presented an SPH scheme 
to model a free-surface incompressible flow which , in analogy with 3D 
Large Eddy Simulations, assumes a \cite{sm63} model for the filtered 
Navier-Stokes equations.

The results of the blob test simulations demonstrate that the instabilities 
leading to the expected cloud disruption can develop only when the SPH
energy equation incorporates the AC term. A particularly interesting 
result is that an appropriate limiting condition must be implemented
 on the AC coefficients $\alpha^C$ in order to avoid an unphysical amount
of heat diffusion, which in turn leads to a cloud disruption which occurs
too early. This limiter has been identified as given by the Prandtl number and,
 for the AC signal velocity adopted here, it severely limits the amplitude of 
the AC coefficients in the regime of strong supersonic flows. 

AC-SPH simulations of the blob test incorporating now the new 
constraint 
support this view, since Fig. \ref{fig:mloss} shows for the new runs a cloud 
mass-loss rate which is in better agreement with the rates obtained from
simulations realized using a completely independent numerical scheme \citep{sp11}. 
However, it must be stressed that the condition (\ref{prac.eq})  has been 
calculated for a perfect monoatomic gas with $\gamma=5/3$, so that a physically
motivated constraint on the artificial heat diffusion of the simulated 
medium should be considered problem dependent.

However, the code has still several problems which render its use problematic 
if the development of hydrodynamic instabilities need to be followed in the 
regime of subsonic flows.  
 The results of the simulations indicate that these 
shortcomings are not due to the AC implementation, but rather are intrinsic 
in the standard formulation with which gradients are calculated in SPH 
and the related errors are subsequently introduced in the momentum equation.
In particular, simulations of the RT test show that increasing the kernel
order alleviates the problems but does not solve them. 
These results are in accordance with recent 
findings \citep{na11,ga12} and clearly demonstrate that for very subsonic flows,
the poor performances of SPH to model hydrodynamic  instabilities  
are strictly connected to the code accuracy in gradient estimates.

The formulation of \cite{ga12}, which has been proposed with the aim of 
calculating SPH gradients with an as high as possible accuracy while keeping the 
benefits of a Lagrangian formulation, 
looks in this aspect very promising and suggests 
further investigations along the numerical approach proposed in this paper.
These would be particularly relevant in the light of the recent results of 
\cite{ba12}, who claim that standard SPH fails  to properly model the regime 
of subsonic turbulence. They reach their conclusions by comparing results 
extracted from simulations of driven subsonic turbulence realized using 
the moving-mesh code AREPO and GADGET SPH. Their results have been criticized 
by \cite{pr12}, for whom the use of a time-dependent AV
 is critical in SPH simulations of subsonic turbulence. 

A re-analysis of the \cite{ba12} simulations 
using the AC-SPH code presented here, augmented with improved gradient 
operators, would then be of fundamental importance to  achieve a deeper 
understanding of the capability  of different numerical methods to model 
 subsonic turbulence.  
The latter being expected to have a significant impact in shaping the 
thermodynamic properties of baryons in cosmological haloes and, subsequently, 
 the process of galaxy formation.


%% References
%%

\end{document}